\documentclass[twocolumn]{aastex631}
\usepackage{booktabs} 
\usepackage{tablefootnote} 
\usepackage{threeparttable}
\usepackage{amsmath}
\usepackage{graphicx} 
\usepackage{float} 
\usepackage{multirow}

\begin{document}

\title{Demystifying Strange \ion{O}{6} Line Widths with Hydrodynamic Simulations 

 }

\author[0009-0009-2215-3941]{Chen Wang}
\affiliation{Department of Physics and Astronomy and Center for Simulational Physics, University of Georgia, Athens, GA, 30602, USA} 

\author[0009-0007-3193-0090]{Eric Goetz}
\email{eric.goetz@uga.edu, chen.wang3@uga.edu, robinl1@uga.edu}
\affiliation{Department of Physics and Astronomy and Center for Simulational Physics, University of Georgia, Athens, GA, 30602, USA} 

\author[0000-0001-5221-0315]{Robin Shelton}
\affiliation{Department of Physics and Astronomy and Center for Simulational Physics, University of Georgia, Athens, GA, 30602, USA} 


\begin{abstract}

We present a survey of \ion{O}{6} {line widths} obtained from hydrodynamic simulations that model the mixing between High Velocity Clouds and the circumgalactic medium and track the {non-equilibrium ionization} populations of the ions. We run 10 simulations with various physical conditions of the clouds and ambient environments, so that our results can be compared to observations of various cloud environments. Synthetic spectra are created for the simulated sight lines that contain \ion{O}{6}. The range of our Doppler broadening parameter, $b$, is from {$\sim$6 km s$^{-1}$ to $\sim$107 km s$^{-1}$}. We calculate the thermal and non-thermal contributions to the $b$ values. Both $b$ and the thermal contribution to $b$ can be substantially less than the CIE value. These narrow line widths are due to the time delay in recombination from \ion{O}{6} to \ion{O}{5} in mixing regions, which results in substantial amounts of \ion{O}{6} at temperatures below the {2.9} $\times$ 10$^{5}$ K CIE temperature. Our results show that non equilibrium collisional ionization and mixing can produce the narrow line widths that are seen in multiple observations.  

\end{abstract}

\section{Introduction} 


The Doppler broadening parameter, $b$, plays an important role in understanding the thermal and kinematic properties of astrophysical environments. 
Wide profiles often appear in systems with active heating or strong dynamical interactions, such as the Circumgalactic Medium (CGM) and Intergalactic Medium (IGM) \citep{Werk2016, Stocke2019, Qu2024}.  High Velocity Clouds (HVCs) interacting with the Galactic halo can also exhibit broad lines, reflecting a combination of thermal motion and turbulence \citep{Lehner2007, Fox2010} 


An interesting puzzle arises: why are some observed line widths broad and {others} extremely narrow, when the atoms, such as \ion{O}{6}, are highly ionized and presumably hot? 
Such narrow line widths are found in the spectra from HVCs \citep{Zech2008}, the {{IGM \citep{Tripp2008, Wakker2009, Savage2014}}} and the CGM of other galaxies \citep{Ahoranta2021, Haislmaier2021, Qu2024}.


To solve this puzzle, we calculate $b$ values from hydrodynamic simulations of HVCs moving through the CGM. 
The simulations track the hydrodynamic evolution as cool cloud material is ablated and hot ambient material is entrained into the cool clouds.  
{Our simulations do not model photoionization but they do track the collisional} ionization and recombination of oxygen in a time dependent manner, determining the non-equilibrium ionization (NEI) population in every ionization level of oxygen.

As a result of NEI and mixing, there are substantial populations of both cool \ion{O}{6} and hot \ion{O}{6} in the clouds. Meanwhile, there are also strong velocity gradients that contribute to the total line width along some sight lines. 

 Our simulations reproduce a wide range of line widths, including narrow values ($b <$ 10 km/s) similar to those observed by {{\citet{Tripp2008} and }} \citet{Savage2014}, {which are noticeably narrower than that of gas at the collisional ionizational equilibrium (CIE) temperature of \ion{O}{6}, and very broad values (b$>$70 km s$^{-1}$), similar to the broad lines in a small number of samples seem in \citet{Savage2014} and \citet{Tripp2008}}. We also calculate the thermal broadening. By removing it from the total line width, we determine the {non-thermal contribution, which is generally associated with {velocity gradients and} turbulent mixing.} {Most of our non-thermal components are narrow, which is similar to the results of \citet{Tripp2008} and \citet{Savage2014}.}

 {In Section~\ref{sec:s2}, we present the simulation code, settings and initial physical conditions. In Section~\ref{sec:s3}, we present our simulations, explain our analyses of the simulations, describe the calculations of the line widths, and present the results. In Section~\ref{sec:s4}, we compare our results to various observations and discuss our physical insights. We }{conclude in Section~\ref{sec:s5}} 

\section{Model} 
\label{sec:s2}

We use FLASH version 4.6.2 \citep{Fryxell2000} to simulate the hydrodynamics of HVCs moving through the ambient environment.  We use FLASH's NEI module to track the time-dependent ionization state populations of the metals.  We also include radiative cooling, calculated using the CIE cooling curve from \citet{Sutherland1993} with [Fe/H] = -0.5. {We do not expect the results to be highly sensitive to the metallicity. \citet{Henley2017} simulated Smith Cloud-like high velocity materials using this metallicity and the associated cooling curve. They also simulated the cloud using solar metallicity and the associated cooling curve. Aside from the different metallicities, the overall characteristics of their simulated clouds are similar to each other.}

We run our simulations in a rectangular domain with lengths 2.4 kpc $\times$ 1.2 kpc $\times$ 10.8 kpc in the $\hat{x}$, $\hat{y}$, and $\hat{z}$ directions, respectively.  The domain is subdivided into 18 blocks, each with a length of 1.2 kpc.  Each of these blocks is initially subdivided into many small cells. The adaptive mesh refinement module divides blocks into smaller blocks in regions where greater spectral resolution is needed. As a result, the minimum cell size is 9.375 pc in each direction. {Furthermore, during the processing of each timestep, each cell is further subdivided into 8 zones in each direction.}

We place the {domain's} origin midway between the lower and upper $x$ boundaries, on the lower $y$ boundary, and 1.2~kpc above the lower $z$ boundary. We place the center of the cloud at the origin. 
Only half of the cloud is within the computational domain. The other half of the cloud is outside the domain, but, owing to the reflection conditions we have imposed at the lower $y$ boundary, is symmetric to the simulated half. 
We allow material to exit the domain through the upper and lower $x$ boundaries, upper $y$ boundary, and upper $z$ boundary. In order to simulate clouds moving quickly through the ambient material, yet avoid having an impractically large domain, we let the cloud be stationary and the ambient gas move past it with a large velocity.
Thus, the simulations use a wind tunnel configuration, with ambient material entering the domain through the lower $z$ boundary. 
At the beginning of each simulation, the cloud is stationary with respect to the domain.

\begin{table*}[t]
    \centering
    \caption{Simulation   {Parameters}}
    \begin{tabular}{lcccccc}
      Simulation & $n(\text{H})_{\text{cl}}$ & $T_{\text{cl}}$ & $n(\text{H})_{\text{am}}$ & $T_{\text{am}}$ & $r_{\text{cl}}$ & velocity\\ 
                 & (cm$^{-3}$) & (K) & (cm$^{-3}$) & (K) & (pc) & (  {km~s$^{-1}$})\\
      \hline           
      Run 1 & 0.4 & 5000 & 0.001 & $2\times10^6$ & 500 & 150 \\ 
      Run 2 & 0.4 & 5000 & 0.001 & $2\times10^6$ & 500 & 100 \\ 
      Run 3 & 0.4 & 5000 & 0.001 & $2\times10^6$ & 300 & 150 \\ 
      Run 4 & 0.4 & 5000 & 0.001 & $2\times10^6$ & 500 & 300 \\ 
      Run 5 & 0.04 & 5000 & 0.0001 & $2\times10^6$ & 500 & 150\\ 
      Run 6 & 0.2 & 5000 & 0.001 & $1\times10^6$ & 500 & 150 \\ 
      Run 7 & 2.0 & 1000 & 0.001 & $2\times10^6$ & 500 & 150 \\ 
      Run 8 & 1.0 & 1000& 0.001 & $1\times10^6$ & 500 & 150 \\ 
      Run 9 & 0.67 & 3000 & 0.001 & $2\times10^6$ & 500 & 300 \\ 
      Run 10 & 0.222 & 9000 & 0.001 & $2\times10^6$ & 500 & 150 \\ 
    \end{tabular}
    \label{tab:params}
\end{table*}

To distinguish between cloud and ambient gas, we initially set each to have a different metallicity.  We start the ambient environment with solar metallicity gas and start the cloud with a 1/1000 of that metallicity.  During post-processing, we reset the initial cloud metallicity to 0.1 solar and the initial ambient metallicity to 0.7 solar, assuming the solar abundances of \citet{Asplund2009}.  Details on the metallicity rescaling process can be found in \citet{Goetz2024}.  

\begin{figure*}[t]
    \centering
    \includegraphics[width=0.85\textwidth]{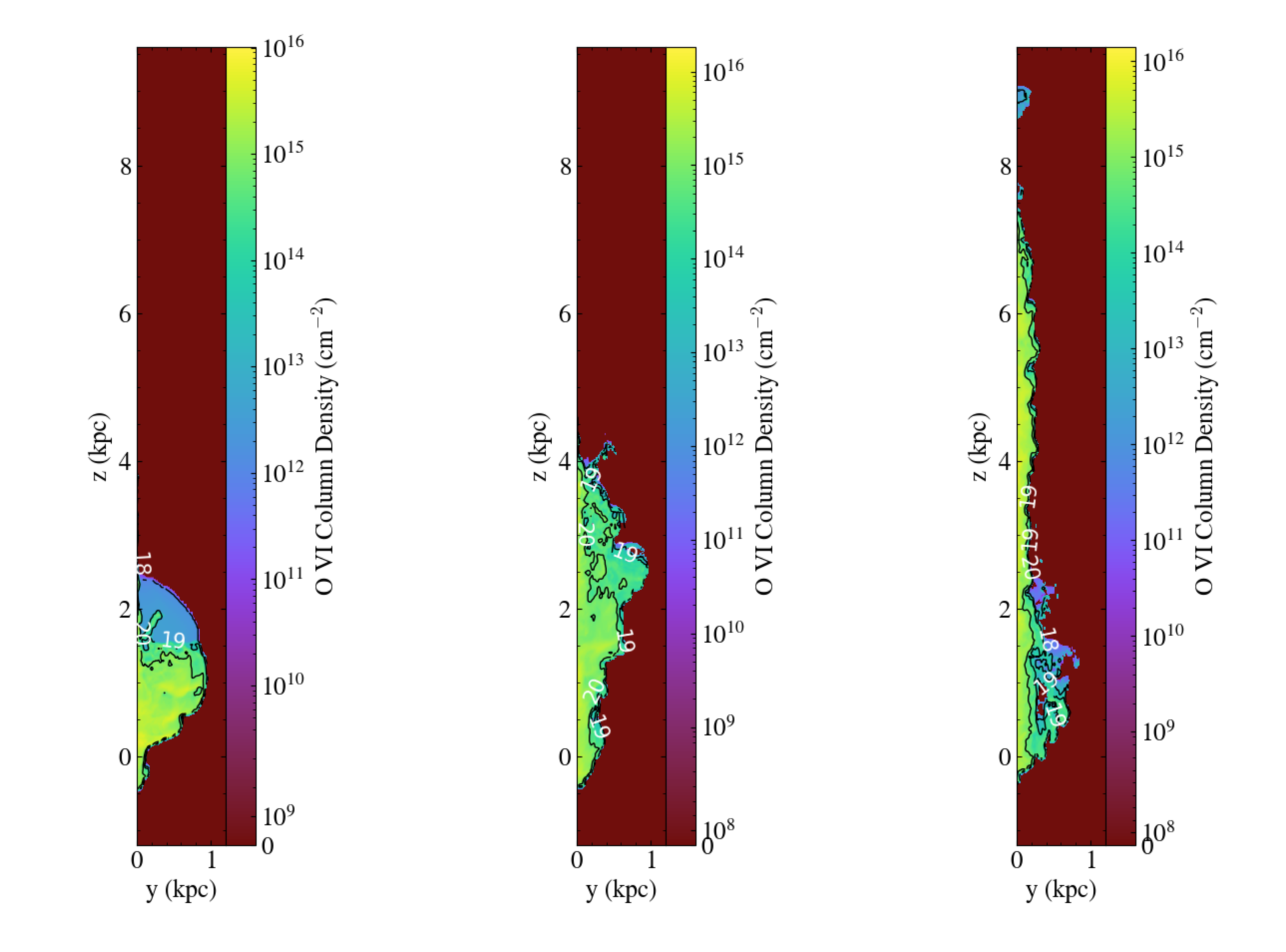}
    \caption{Column density map of \ion{O}{6} at 100, 150 and 200 Myrs for Run 1. The color bar presents the value of the \ion{O}{6} column density, N(\ion{O}{6}). The contour lines and their labeled values represent the {log of the} hydrogen column density. {For each of the three panels, the innermost contour line represents a value of 20 for the $\log$[N(H)], the next contour line represents a value of 19 for the $\log$[N(H)] and the outermost contour line represents a value of 18 for the $\log$[N(H)].}}
    \label{fig:F1}
\end{figure*} 

\begin{figure*}[t] 
    \centering
    \includegraphics[width=0.85\textwidth]{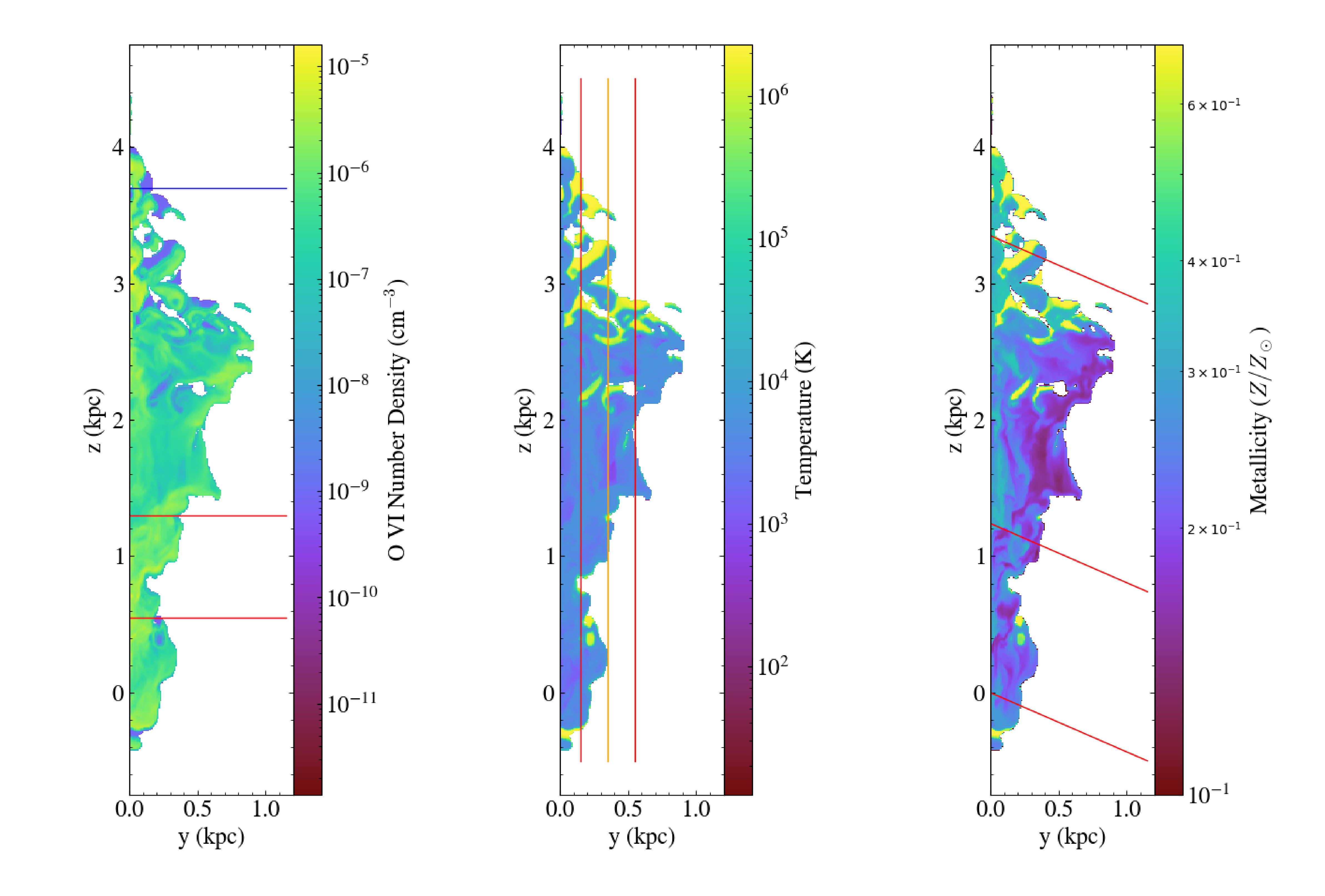}
    \caption{{Left Panel: \ion{O}{6} number density in the cloud, along a slice through the domain in the x=0 plane at 150 Myrs from Run 1. The color bar displays the \ion{O}{6} number density. Middle Panel: Temperature in the cloud, along a slice through the domain in the x=0 plane at 150 Myrs from Run 1. The color bar displays the temperature. Right Panel: Metallicity in the cloud, along a slice through the domain in the x=0 plane at 150 Myrs from Run 1. The color bar displays the metallicity. 
    {The figures do not show background material that is not moving with the cloud's velocity.}
    The red, blue and orange lines are the simulated sight lines. The orange line in the middle panel is the sight line in Figure~\ref{fig:F3}, and the blue line in the left panel is the sight line in Figure~\ref{fig:F4}. }}
    \label{fig:F2}
\end{figure*} 

\begin{figure*}[t]
    \centering
    \includegraphics[width=0.85\textwidth]{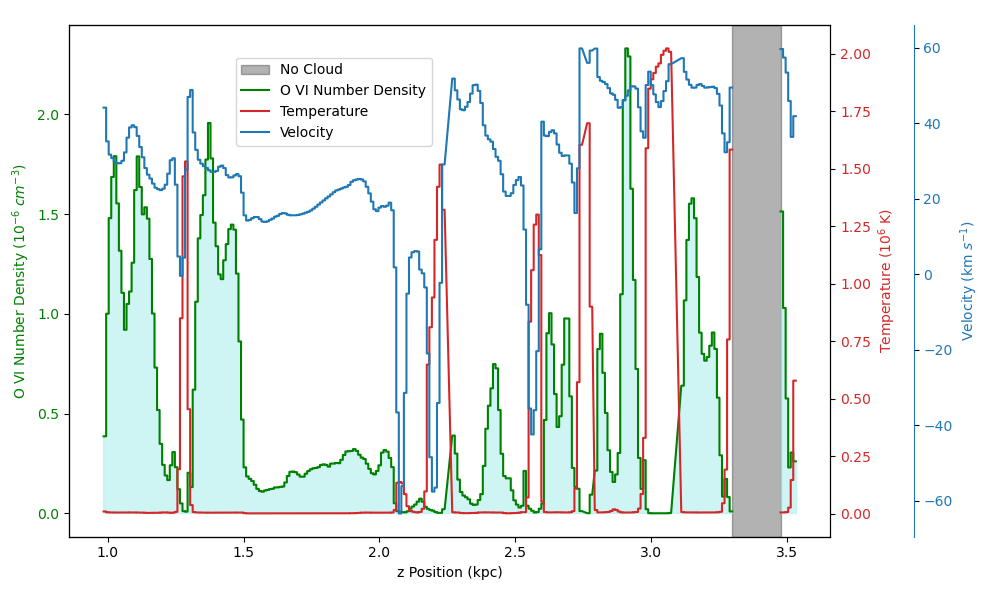}
    \caption{{Temperature, \ion{O}{6} number density and velocity along the sight line marked in orange in the middle panel of Figure~\ref{fig:F2}. The plotted velocity is in the $\hat{z}$ direction and is relative to the domain's reference frame, in which the cloud's initial velocity was 0 km s$^{-1}$. The gray band marks a region that is filled with ambient material.}}
    \label{fig:F3}
\end{figure*}   

\begin{figure*}[t]
    \centering
    \includegraphics[width=0.85\textwidth]{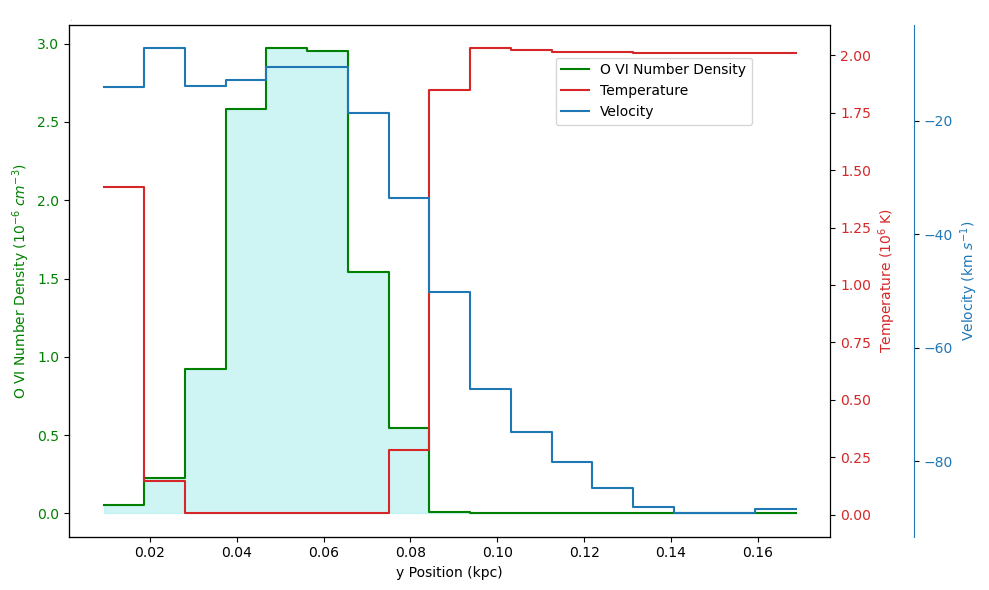}
    \caption{{Temperature, \ion{O}{6} number density and velocity along the sight line marked in blue in the left panel of Figure~\ref{fig:F2}. The plotted velocity is in the $\hat{y}$ direction.}}
    \label{fig:F4}
\end{figure*}  

\begin{figure*}[] 
    \centering
    \includegraphics[width=0.85\textwidth]{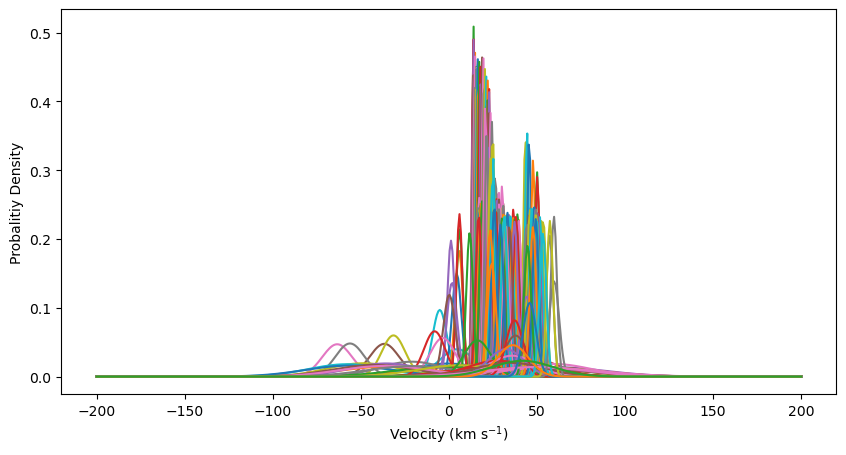}
    \caption{Maxwell-Boltzmann distributions for all the cells along a sight line, calculated from their temperatures and bulk velocities. Each curve is for a single cell. {The location of this sight line is marked in orange in the second panel of Figure~\ref{fig:F2}. In this plot, we can see two obvious clusters of cool gas with one centered around 20 km s$^{-1}$ and the other one centered around 45 km s$^{-1}$ The curves have various colors in order to separate them visually.} }
    \label{fig:F5}
\end{figure*} 

\begin{figure*}[]
    \centering
    \includegraphics[width=0.85\textwidth]{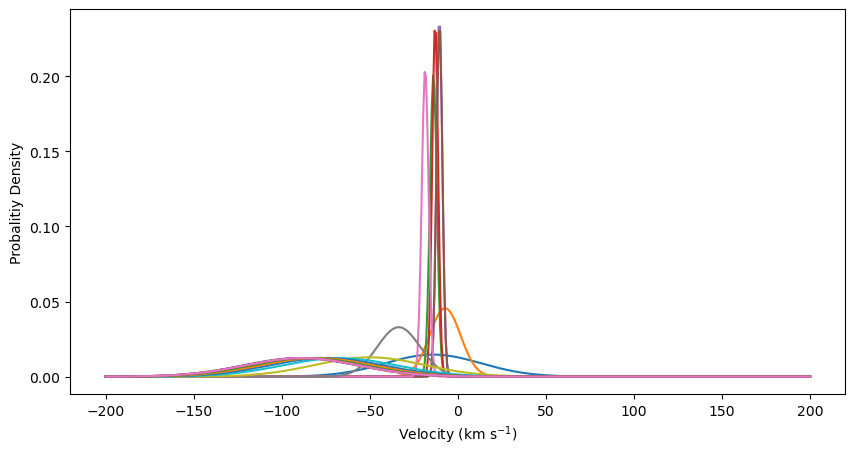}
    \caption{Maxwell-Boltzmann distributions for all the cells along a sight line, calculated from their temperatures and bulk velocities. Each curve is for a single cell. {The location of this sight line is marked in blue in the first panel of Figure~\ref{fig:F2}. The curves have various colors in order to separate them visually.}}
    \label{fig:F6}
\end{figure*} 

\begin{figure*}[t]
    \centering
    \includegraphics[width=0.85\textwidth]{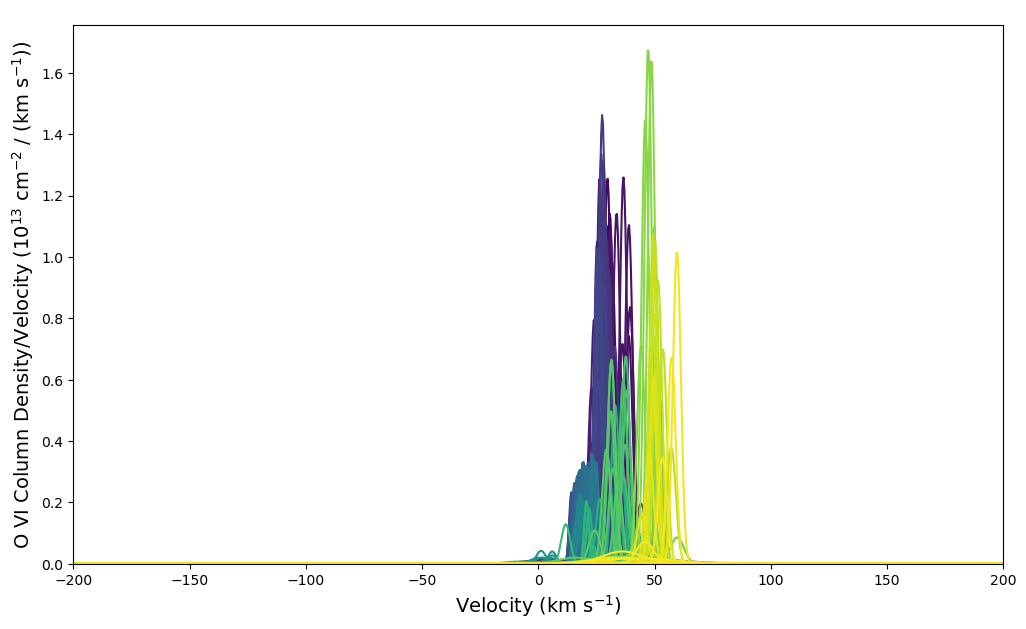}
    \caption{{\ion{O}{6} distributions for individual cells along the sight line marked in orange in the middle panel of Figure~\ref{fig:F2}.} {Each curve is for a single cell. The curves have various colors in order to separate them visually. }}
    \label{fig:F7}
\end{figure*}  

\begin{figure*}[t]
    \centering
    \includegraphics[width=0.85\textwidth]{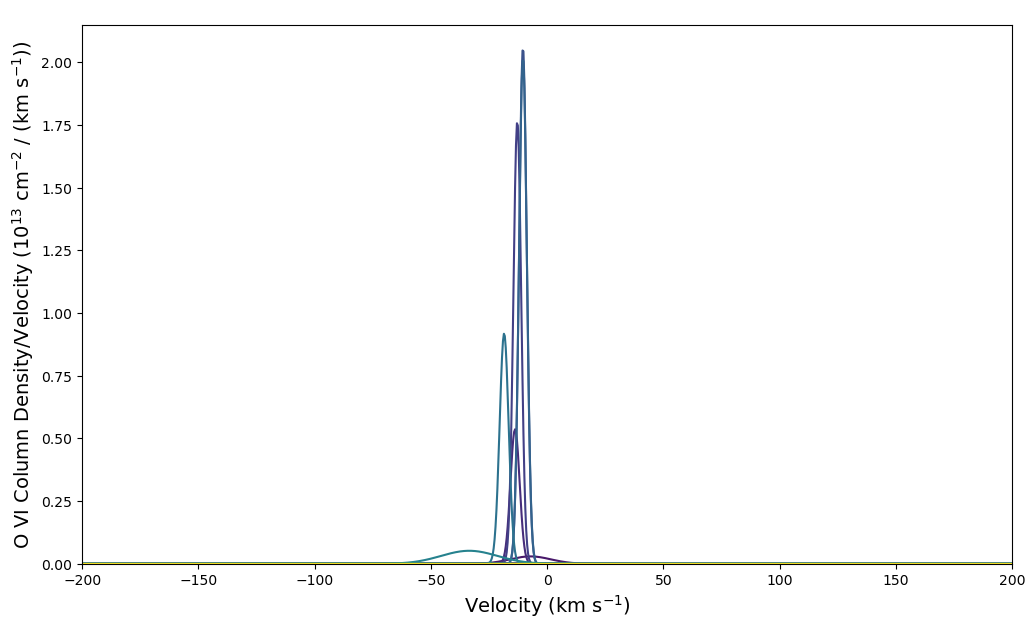}
    \caption{{\ion{O}{6} distributions for individual cells along the sight line marked in blue in the first panel of Figure~\ref{fig:F2}. Each curve is for a single cell. The curves have various colors in order to separate them visually. }}
    \label{fig:F8}
\end{figure*} 

\begin{figure*}[] 
    \centering
    \includegraphics[width=0.85\textwidth]{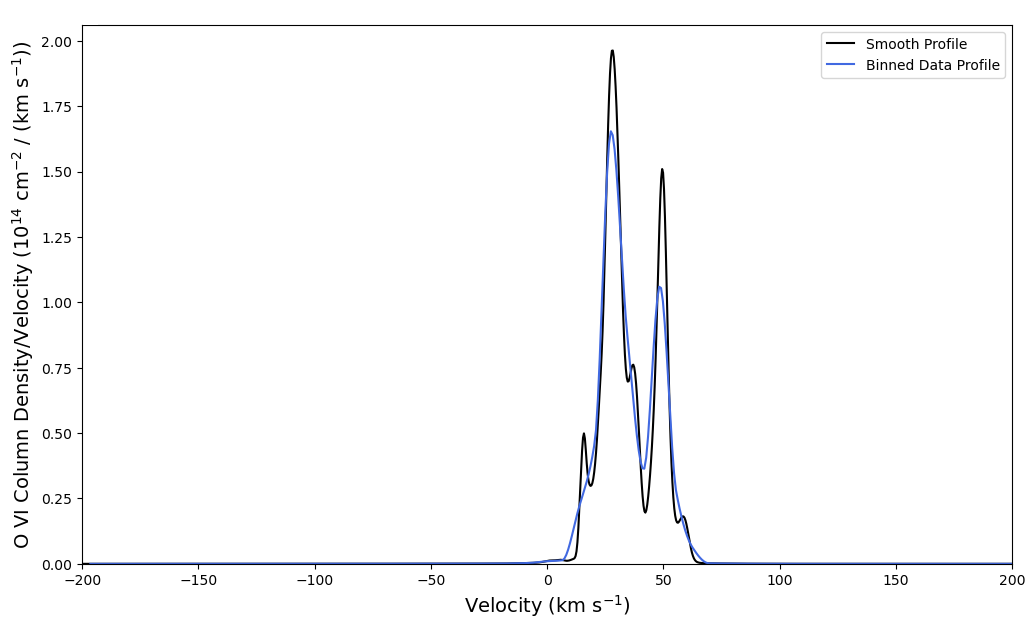}
    \caption{{Plot of \ion{O}{6} column density vs. velocity for the orange sight line in the second panel of Figure~\ref{fig:F2}. The black curve is the direct sum of the curves in Figure~\ref{fig:F7}. It has been binned over 7 km s$^{-1}$ to make the blue curve.} }
    \label{fig:F9}
\end{figure*} 

\begin{figure*}[t]
    \centering
    \includegraphics[width=0.85\textwidth]{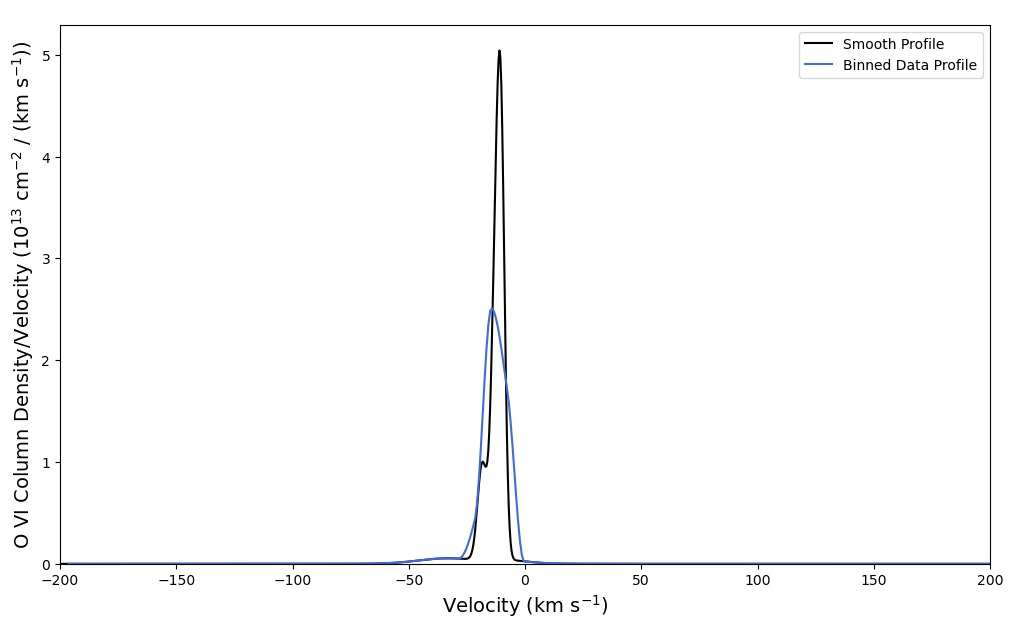}
    \caption{{Plot of \ion{O}{6} column density vs. velocity for the blue sight line in the first panel of Figure~\ref{fig:F2}. The black curve is the direct sum of the curves in Figure~\ref{fig:F8}. It has been binned over 7 km s$^{-1}$ to make the blue curve.}}
    \label{fig:F10}
\end{figure*} 

\begin{figure*}[t]
    \centering
    \includegraphics[width=0.85\textwidth]{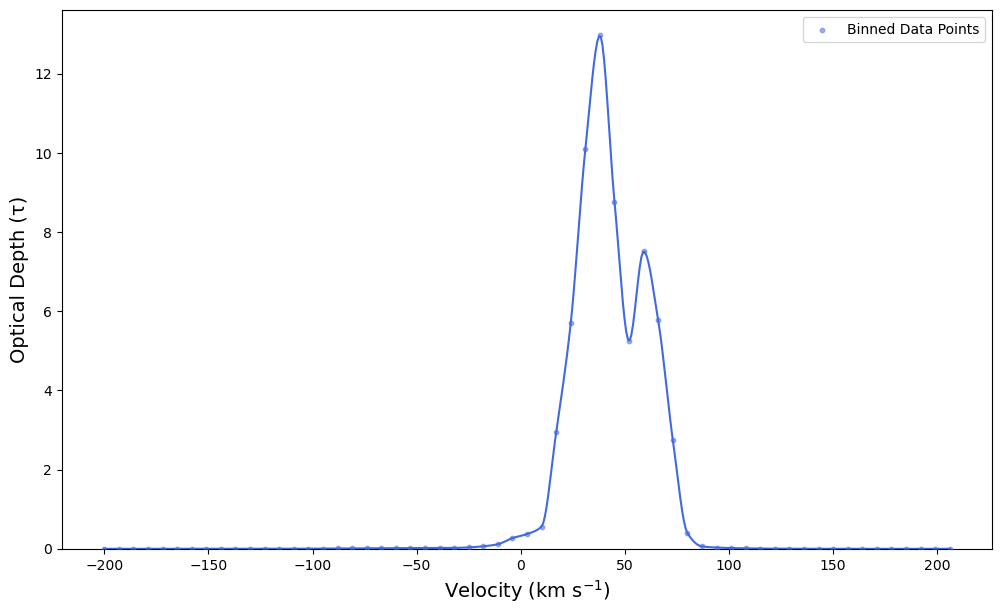}
    \caption{{Plot of optical depth, $\tau$ vs. velocity from Trident results for the orange sight line in the second panel of Figure~\ref{fig:F2}. Trident models the effect of LSF and this is why the shape of the curve here is slightly different from the one in Figure~\ref{fig:F9}.}}
    \label{fig:F11}
\end{figure*} 

\begin{figure*}[t]
    \centering
    \includegraphics[width=0.85\textwidth]{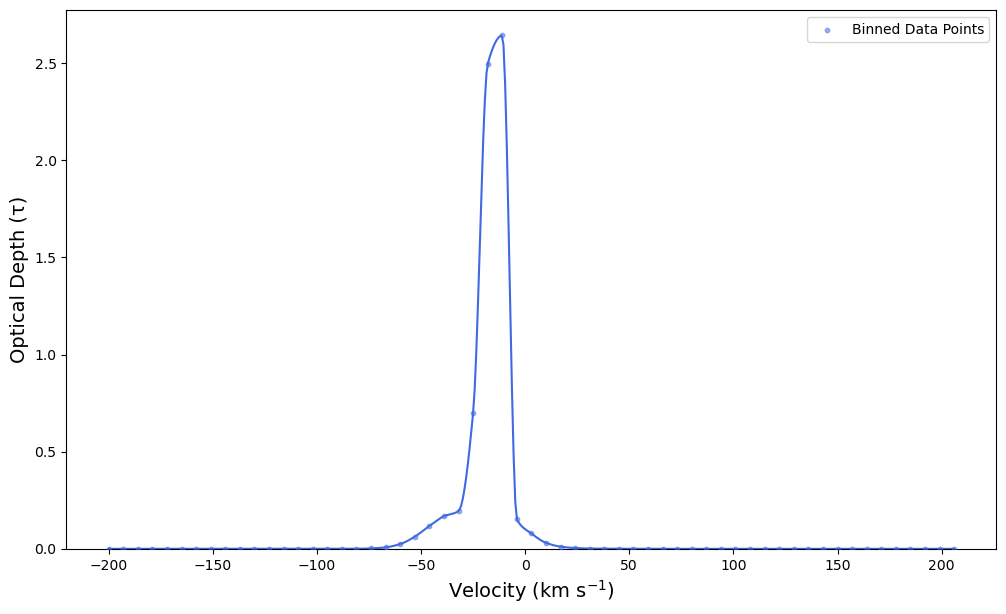}
    \caption{{Plot of optical depth, $\tau$ vs. velocity from Trident results for the blue sight line in the first panel of Figure~\ref{fig:F2}. Trident models the LSF effect and this is why the shape of the curve here is slightly different from the one in Figure~\ref{fig:F10}.}}
    \label{fig:F12}
\end{figure*}

\begin{figure*}[t]
    \centering
    \includegraphics[width=0.85\textwidth]{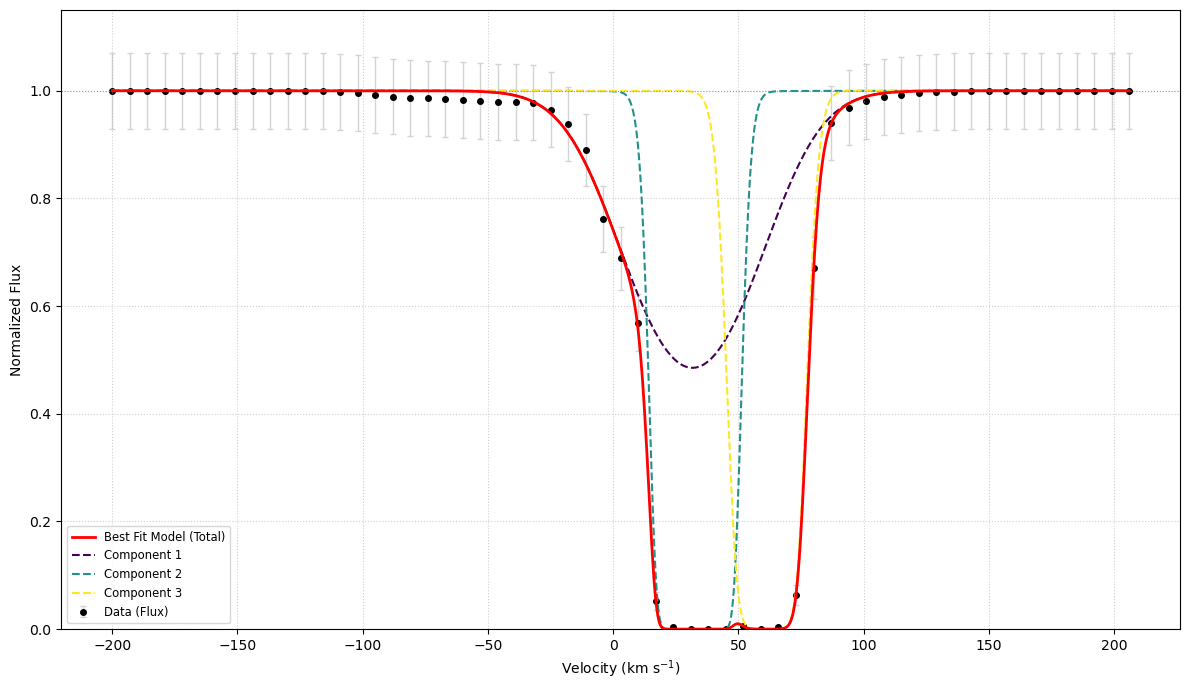}
    \caption{{Final \ion{O}{6} absorption spectrum from the sight line that is marked in orange color in Figure~\ref{fig:F2}. }}
    \label{fig:F13}
\end{figure*}

\begin{figure*}[t]
    \centering
    \includegraphics[width=0.85\textwidth]{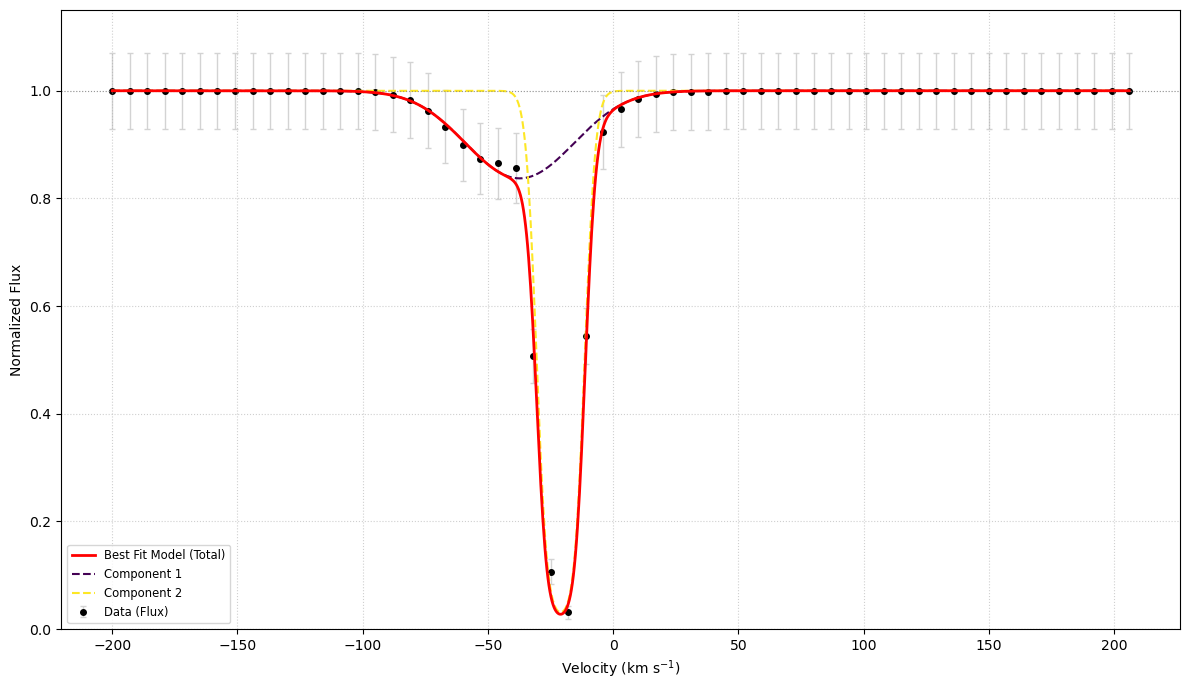}
    \caption{{Final \ion{O}{6} absorption spectrum from the sight line that is marked in blue color in Figure~\ref{fig:F2}.}}
    \label{fig:F14}
\end{figure*}

In order to cover a wide range of HVC parameters, we run ten simulations with various cloud and ambient densities and temperatures, cloud radii, and wind speeds.  The parameters for these simulations are given in Table~\ref{tab:params}.
The cloud density and temperature given in Table \ref{tab:params} are the values at the center of the cloud.  The cloud density decreases with distance from the center of the cloud until it reaches the ambient values.  The shape of the density profile is given by the hyperbolic tangent function described in \citet{Gritton2014}, with a scale length of 50 pc.  The cloud and ambient environment start in pressure equilibrium.  Therefore, the cloud temperature rises towards the edge of the cloud until it reaches the ambient temperature, governed by a complementary hyperbolic tangent function.  At the beginning of the simulation, the metallicity and velocity transition from the cloud's values to the ambient values at the locations where the hydrogen number density is $0.99 n_{\text{am}}+0.01 n_{\text{cl}}$.  Here, $n_{\text{am}}$ is the initial ambient hydrogen number density and $n_{\text{cl}}$ is the initial cloud hydrogen number density.

During the simulations, the cloud material mixes together with the ambient material. In the real world, observers would distinguish the cloud material from the ambient material by differences in velocity. Likewise, we follow similar logic. {We define our own velocity criterion; the details} can be found in \citep{Goetz2024}.

\section{Results} 
\label{sec:s3}

{Figure~\ref{fig:F1} shows one of our simulations at three epochs (t=100, 150, 200 Myrs).} {We invented sight lines that pass in $\hat{y}$, $\hat{z}$ or oblique directions though the cloud.    
Figure~\ref{fig:F2} shows the sight lines superimposed on {\ion{O}{6} number density, temperature and metallicity} maps of Run 1 at 150 Myrs.} 

{Many sight lines transect \ion{O}{6} at a variety of temperatures. Sometimes multiple cloudlets have very similar velocities. These phenomena occur along the sight line highlighted in orange in the middle panel of Figure~\ref{fig:F2}, whose temperature, \ion{O}{6} number density, and velocity are shown in Figure~\ref{fig:F3}. There are several separate cool clumps of \ion{O}{6}-rich gas along the sight line. Some are at velocities of $\sim$50 km s$^{-1}$ to $\sim$60 km s$^{-1}$ and some are at velocities of $\sim$30 km s$^{-1}$. Later in this section, we show how this gas would appear in a synthetic observation; the multiple distinct clumps at velocities of $\sim$50 km s$^{-1}$ to $\sim$60 km s$^{-1}$ would not be observationally separable. Same for the cloudlets of $\sim$30 km s$^{-1}$.}  

{Additionally, single large gas structures can show smooth changes in velocity from one side to the other; this velocity gradient may come from large-scale motions, such as flows or shearing within the gas. An example of such gradients is shown in Figure~\ref{fig:F4}. It is for the sight line highlighted in blue in Figure~\ref{fig:F2}. Note the large velocity gradient from $\sim$0.06 to $\sim$0.15 kpc.} 

{We calculated the spectra on these two sight lines, the seven other sight lines shown in Figure~\ref{fig:F2} for Run 1 at 150 Myrs, nine similar sight lines through the domain at 100 Myrs, nine similar sight lines at 200 Myrs, and nine similar sight lines through each of Run 2 through Run 10 at each of these epochs. }
In order to calculate the synthetic spectra, we first assumed that the gas in every cell along a line of sight has a Maxwell-Boltzmann distribution of velocities, centered on the cell's bulk velocity. We calculated the Maxwell-Boltzmann distribution along the line of sight from the formula

\begin{equation}
f(v) = \frac{1}{\sigma \sqrt{2\pi}} e^{-\frac{(v - \mu)^2}{2\sigma^2}},
\end{equation} 
where $\mu$ is the cell's bulk velocity along the line of sight and $\sigma$ is the one-dimensional velocity dispersion of gas in the cell due to thermal broadening. The one-dimensional velocity dispersion is calculated by 
\begin{equation}
\sigma = \sqrt{\frac{kT}{m}},
\end{equation}
where $k$ is the Boltzmann constant, $T$ is the temperature of the gas in the cell and $m$ is the mass of an oxygen atom. 


{{Figure~\ref{fig:F5}} shows examples of the one-dimensional Maxwell-Boltzmann distributions for all of the cells along the sight line in the $\hat{z}$ direction marked in orange in the second panel of Figure~\ref{fig:F2} for Run 1 at 150 Myrs. {Figure~\ref{fig:F6}}  shows the same thing, but for the sight line in the $\hat{y}$ direction marked in blue in the first panel of Figure~\ref{fig:F2}.} 
{These plots show the effects of thermal broadening and the variation in bulk velocities. }

{Considering that the {bulk velocity, thermal broadening, and number of \ion{O}{6} ions} in any given cell need not match {those} of the other cells along the line of sight, we created a column density array in which the contributions of each cell could be combined. The array spans the velocity range expected for the line of sight and is segmented into many “bins”, each of which spans 1 km s$^{-1}$. 
For any given cell, the fraction of material within any given velocity bin is the integral of $f(v) dv$ between the bin's lower and upper velocity boundaries. 
We multiplied these fractions by the \ion{O}{6} column density of the cell to get each cell's contribution of \ion{O}{6} within each velocity bin. Thus, every single cell has an array of the amount of \ion{O}{6} with respect to the center value of the velocity bin. Plots of these arrays represent the contribution of \ion{O}{6} gas from individual cells along the sight line, after considering the \ion{O}{6} number density, thermal broadening effect and bulk velocity.} 

{Figure~\ref{fig:F7} and Figure~\ref{fig:F8} show examples from the sight lines in the $\hat{z}$ and $\hat{y}$ directions  that are highlighted in Figure~\ref{fig:F2}.}  
{Figure~\ref{fig:F7} is a column density-weighted version of Figure~\ref{fig:F5}. In general, the narrow curves in Figures~\ref{fig:F5}  and \ref{fig:F7} come from cells in the cool clumps shown in Figure~\ref{fig:F3}. Many of these curves overlap each other to form two obvious conglomerations.}  

{Likewise, Figure~\ref{fig:F8} is a column density-weighted version of Figure~\ref{fig:F6}. The narrow curves in Figure~\ref{fig:F8} come from the large \ion{O}{6} clump in Figure~\ref{fig:F4}. Their central velocities range from $\sim$-22 km s$^{-1}$ to $\sim$-10 km s$^{-1}$. This gradient is due to the velocity gradient on the right side of the \ion{O}{6} clump in Figure~\ref{fig:F4}. In addition, there are two shallow, broad curves from warm gas at temperature of $\sim$2 to $\sim$3 $\times$ 10$^{5}$~K. In Figure~\ref{fig:F6}, there are also several very broad, shallow curves at more negative velocities. These are due to swept up hot, diffuse ambient material. They contain little \ion{O}{6}, and so are imperceptible in Figure~\ref{fig:F8}.
}


{We summed the distributions in Figure~\ref{fig:F7} to create the spectra of N(\ion{O}{6}) vs. velocity in Figure~\ref{fig:F9}. (The black curve is the direct sum; the blue curve is the binned version of the black curve.) Here, the two conglomerations in Figures~\ref{fig:F5} and \ref{fig:F7}  make the two obvious features in Figure~\ref{fig:F9}.
This gas comes from the cool cloudlets transected in Figure~\ref{fig:F3}. The velocity dispersion between the constituent curves in each conglomeration in Figure~\ref{fig:F7} contributed to the overall breadth of the resulting features in Figure~\ref{fig:F9}. We also found a correspondence between clusters of cool cells in Figure~\ref{fig:F6}, ~\ref{fig:F8} and the narrow feature in the summed spectra (Figure~\ref{fig:F10}). } 

{In a real observation, astronomers will extract the summed spectra and then try to disentangle the velocity components. This is often done with absorption spectroscopy. We used Trident \citep{Hummels2017}, to simulate an absorption spectra for sight lines through our domain. 
The instrument we assumed when running Trident to simulate the observational sight line is the Space Telescope Imaging Spectrograph E140M. We chose this instrument because it has good resolution (~7 km s$^{-1}$) (\citet{STIS2024}, and see also \citet{Linsky2025}) and a large amount of \ion{O}{6} observational data has been obtained with it,  (e.g. \citet{Tripp2008}). One issue is that \ion{O}{6}'s rest frame wavelengths are outside the bandpass of the E140M grating. Thus, we shifted \ion{O}{6}'s wavelengths into the E140M grating's bandpass by adjusting the redshift of the simulated material.  We used a value of z=0.3 and after this shift our \ion{O}{6}'s wavelengths are within the range from 1300 \AA\ to 1350 \AA\ where the E140M grating's sensitivity peaks. In addition, a survey done by \citet{Tripp2008} found a median redshift of z=0.217 for intervening \ion{O}{6} systems and a median redshift of z=0.267 for proximate absorbers. Considering their redshift values and the E140M sensitivity range, z=0.3 is a reasonable choice. We removed this redshift later in our analysis in order to produce the final spectra.}

{In our Trident analysis, we incorporated the tabulated line spread function (LSF) of the STIS E140M provided by the Space Telescope Science Institute into Trident. We used the 0.06$''$ × 0.2$''$ aperture, which was the most commonly used aperture in the \citet{Tripp2008} survey. Combining all of these effects, the resolution of our final spectra is close to 7 km s $^{-1}$, which is the resolution of the E140M grating. We did not add noise. Trial runs with noise, using exactly the same S/N ratio and other settings, resulted in a wide range of seemingly random b values. The variation among the trials could be $\sim$50$\%$ at most.}

{We used Trident to calculate the optical depth of the \ion{O}{6} material, $\tau$, and make a plot of it versus velocity {(Figure~\ref{fig:F11})}.  Compared with {the blue curve in Figure~\ref{fig:F9}} made directly from our FLASH data, the general shape and structure of the Trident plot are preserved while the instrumental LSF effect from Trident changes the shapes and broadens the line widths in {Figure~\ref{fig:F11}}.}

{We then plotted the final spectra of normalized flux vs. velocity and used a Voigt profile in the Python library, SciPy, to identify and fit individual components. {Figure~\ref{fig:F13} and Figure~\ref{fig:F14}} show the absorption spectra from the two sight lines marked in orange and blue in Figure~\ref{fig:F2}. In {Figure~\ref{fig:F13}}, the two deep components are identified; they clearly correspond to the two obvious peaks plotted in {Figures~\ref{fig:F7}, \ref{fig:F9} and \ref{fig:F11}.} A third component which has a wide width is also identified. {This component comes from mixed gas from multiple locations and velocity centroids, plus some of the cool gas.}
Since this sight line is in the $\hat{z}$ direction, it intersects more cloud material than the sight lines in $\hat{y}$ and oblique directions, as shown in Figure~\ref{fig:F2}, resulting in more spectral structures. In {Figure~\ref{fig:F14}}, one deep, narrow component and one shallow, broad component are identified. Similarly, the identified components in {Figure~\ref{fig:F14}} correspond to the {one broad and one narrow feature} in {Figures~\ref{fig:F10} and ~\ref{fig:F12}}. The broad component can be explained by the obvious hot gas which is shown in Figure~\ref{fig:F2}. 
}


{The fitting results also include the estimated velocity centers, $v_{c}$, Doppler broadening parameters, $b$, and $\log N(O~VI)$ for individual components.
Since $b$ embodies the effects of both thermal broadening and velocity gradients, we separately estimated the contribution from thermal broadening, $b_{T}$ with the following method. We first calculated the velocity range for each component as ($v_{c}$ - 2.5 $\sigma_{o}$, $v_{c}$ + 2.5 $\sigma_{o}$), where $\sigma_{o}$ is the standard deviation of the observed line profile for each component. Then we identified the cells within this range along the sight line in our simulation domain. Next, for each individual velocity component, we calculated the weighted average temperature of \ion{O}{6} in these cells.  Except when dealing with broad, relatively shallow components that overlap other components, this weighted average temperature, $T_w$, is calculated from}
{
\begin{equation}
T_w = \frac{\sum_i N(\text{\ion{O}{6}})_i \, T_i}{\sum_i N(\text{\ion{O}{6}})_i},  
\end{equation} 
}
{where $i$ refers to any given cell that is within the velocity range along the chosen sight line, $T_i$ is the temperature of that cell, and $N(\text{\ion{O}{6}})_i$ is the column density of O~VI of that cell.} 

{The calculation of $T_w$ for any given component uses the cells from its specified velocity range. Consequently, for components that overlap, the cells in the shared velocity region would contribute to the calculation for all involved components.
However, the shared cells could lead to an underestimate or overestimate of each component's temperature. This is a bigger problem for the calculation of $T_w$ of broad, shallow components from hot, diffuse gas, like Component 1 in {Figure~\ref{fig:F13}}, than the calculation of $T_w$ of narrow, deep components from cool, dense gas. To avoid this issue for hot components, we only use the cells outside the velocity ranges of the other components when we calculate the values of $T_w$ for the broad, relatively shallow components.}  

{To illustrate why this exclusion is necessary for the broad components, consider the spectrum shown in Figure~\ref{fig:F13}. The fitted broad component marked with a purple dashed line is Component 1. It is overlapped by the fitted narrow components marked with aqua and yellow dashed lines (Components 2 and 3). The depth of Component 2’s absorption exceeds that of Component 1’s absorption across almost the entire width of Component 2. Considering that the majority of the gas within the overlapped velocity region is from the cool, narrow Component 2 and considering that the average temperature of the cells at Component 2’s velocity is much smaller than that of Component 1 outside Component 2’s range, therefore including the overlapped cells in the Component 1 temperature calculation would lower the resulting temperature significantly. Excluding the overlapped cells, the calculated value of $b_T$ for Component 1 is 16.7 km s$^{-1}$. Including the overlapped cells, the calculated value of $b_T$ is 3.6 km s$^{-1}$, a value that is very similar to that of Component 2 (3.4 km s$^{-1}$). We have repeated such comparisons for all of the broad components in the Run 1 sight lines. We found that allowing such narrow components to contaminate the calculations of $b_T$ of the broad components sways the result dramatically. For this reason, we excluded the overlapped regions from the calculations of $b_T$ for the broad components.} 

{However, we did not exclude the overlapped regions when we calculated $b_T$ for the narrow components. One reason for doing this is that the velocity ranges of the narrow components are usually completely overlapped by the broad ones. Another reason is that the broad components contribute very little to the column density found in the overlapped velocity regions, while the narrow components contribute most of the material.} 

{After determining the values of $T_w$, we then calculate the thermal contribution, $b_{T}$, to the values of $b$, as}
{
\begin{equation}
b_{T} = \sqrt{\frac{2kT_{w}}{m_{o}}}, 
\end{equation} } 
{
where $k$ is the Boltzmann constant and $m_{o}$ is the mass of an oxygen atom.}

{By subtracting $b_{T}$ from $b$ in quadrature, we obtained the broadening parameter associated with the velocity gradients,  $b_{nt}$, i.e.,} 
{
\begin{equation}
b_{nt} = \sqrt{b^2-b_{T}^2}.  
\end{equation} 
}

{The values of $v_{c}$, $b$, $b_{T}$, $b_{nt}$, $\log N(O~VI)$ and the total O~VI column density directly calculated from our runs, N(\ion{O}{6})$_{sim}$, for 27 simulated sight lines through Run 1 are tabulated in Table~\ref{tab:s1100}, \ref{tab:s1150} and \ref{tab:s1200}. The values for Run 2 to 10 are tabulated in the Appendix.}

\begin{table*}[!ht]
\caption{Run 1 at 100 Myrs}
\hspace{-1.2cm}
\label{tab:my_complex_table_1kms}
\begin{tabular}{l l|*{5}{c}|c}
\hline
\multicolumn{2}{c|}{\multirow{2}{*}{Sight Lines}} & \multicolumn{5}{c|}{Results} & \multirow{2}{*}{$\log N(O~VI)_{sim}$} \\ 
\cline{3-7}
& & v$_{c}$ (km s$^{-1}$) & $b$ (km s$^{-1}$) & $b_T$ (km s$^{-1}$) & $b_{nt}$ (km s$^{-1}$) & $\log N(O~VI)$ \\ 
\hline
y & y1 
& -21.8$\pm$0.2 & 8.6$\pm$0.2 & 2.2 & 8.3 & 13.87$\pm$0.01 & 15.18 \\ 
&  
& -1.4$\pm$0.1 & 13.4$\pm$0.2 & 2.0 & 13.2 & 14.65$\pm$0.01 &  \\ 
& y2
& -3.1$\pm$0.1 & 15.9$\pm$0.2 & 2.4 & 15.7 & 14.70$\pm$0.01 & 14.96 \\ 
& y3 
& -41.8$\pm$6 & 74.6$\pm$12.4 & 27.0 & 69.5 & 14.21$\pm$0.08 & 14.71 \\ 
& 
& -41.2$\pm$0.9 & 16.0$\pm$2.3 & 7.9 & 13.9 & 14.17$\pm$0.07 &  \\ 
& 
& -4.7$\pm$0.8 & 6.4$\pm$1.4 & 3.8 & 5.1 & 13.94$\pm$0.07 &  \\ 
\hline
z & z1 
& -47.6$\pm$0.5 & 14.5$\pm$1.9 & 4.0 & 13.9 & 15.55$\pm$0.28 & 15.70 \\ 
& 
& -13.1$\pm$2.1 & 71.1$\pm$2.7 & 24.1 & 66.9 & 14.73$\pm$0.03 &  \\ 
& 
& 7.4$\pm$0.5 & 13.1$\pm$1.4 & 2.9 & 12.8 & 14.81$\pm$0.10 &  \\ 
& z2 
& -57.8$\pm$1.0 & 9.7$\pm$1.7 & 4.9 & 8.4 & 13.66$\pm$0.08 & 15.28 \\ 
&  
& 1.1$\pm$0.6 & 29.5$\pm$1.4 & 2.8 & 29.4 & 15.11$\pm$0.05 & \\
& z3 
&  34.6$\pm$0.1 & 16.8$\pm$0.4 & 6.7 & 15.4 & 15.01$\pm$0.03 & 15.09 \\
\hline
oblique & oblique1 
& -2.0$\pm$0.1 & 8.7$\pm$0.1 & 2.9 & 8.2 & 14.17$\pm$0.00 & 14.70 \\ 
& 
& 6.0$\pm$0.1 & 21.4$\pm$0.1 & 3.7 & 21.1 & 13.31$\pm$0.00 & \\
& oblique2 
& -70.2$\pm$8.9 & 93.0$\pm$21.7 & 20.5 & 90.7 & 13.82$\pm$0.10  & 14.93 \\ 
& 
& -22.9$\pm$0.4 & 14.7$\pm$0.7 & 2.5 & 14.5 & 14.25$\pm$0.02 & \\ 
& 
& 3.7$\pm$0.2 & 8.8$\pm$0.4 & 2.2 & 8.5 & 14.19$\pm$0.03 & \\
& oblique3 
& 11.4$\pm$1.1 & 22.0$\pm$3.0 & 4.1 & 21.6 & 14.79$\pm$0.09 & 14.95 \\ 
& 
& 33.0$\pm$0.1 & 82.6$\pm$17.9 & 13.4 & 81.5 & 14.27$\pm$0.11 & \\
\hline 
\multicolumn{8}{@{}l}{\parbox{0.9\textwidth}{\small\vspace{1mm}{Note: {The velocity center ($v_{c}$), Doppler broadening parameter ($b$), thermal contribution to $b$ ($b_T$), remaining contribution to b ($b_{nt}$), and log of the \ion{O}{6} column density ($\log N(O~VI)$) for each fitted velocity component along each sampled sight line, as well as the log of the \ion{O}{6} column density along each sight line ($\log N(O~VI)_{sim}$). The values of $\log N(O~VI)$ for each velocity component along each sight line were found from the Trident fitting, while each sightline’s value of $\log N(O~VI)_{sim}$ was calculated directly from the simulational domain.
Because $\log N(O~VI)_{sim}$ and $\log N(O~VI)$ were found by different methodologies, the value of $N(O~VI)_{sim}$ for any given sight line need not equal the sum of $ N(O~VI)$ for the components along that sight line. See text for explanations of how these quantities were determined.} Results for sight lines through Run 1 at 100 Myrs. y1, y2, and y3 refer to three sight lines through the clouds in the $\hat{y}$ direction at low, middle, and high values of z. Likewise for the oblique sight lines. z1, z2, and z3 refer to three sight lines through the clouds in the $\hat{z}$ direction at low, middle, and high values of y. For some sight lines (e.g., y1 and y3), there are multiple detected velocity components. }}} \\
\end{tabular} 
\label{tab:s1100} 

\end{table*}

\begin{table*}[!ht]
\caption{Run 1 at 150 Myrs} 
\hspace{-1.2cm}
\label{tab:my_complex_table_1kms_150Myr} 
\begin{tabular}{l l|*{5}{c}|c}
\hline
\multicolumn{2}{c|}{\multirow{2}{*}{Sight Lines}} & \multicolumn{5}{c|}{Results} & \multirow{2}{*}{$\log N(O~VI)_{sim}$} \\ 
\cline{3-7}
& & v$_{c}$ (km s$^{-1}$) & $b$ (km s$^{-1}$) & $b_T$ (km s$^{-1}$) & $b_{nt}$ (km s$^{-1}$) & $\log N(O~VI)$ \\ 
\hline
y & y1 
& -21.9$\pm$0.1 & 8.3$\pm$0.2 & 5.8 & 5.9 & 13.25$\pm$0.01 & 14.96 \\ 
& 
& -2.2$\pm$0.1 & 9.4$\pm$0.1 & 2.1 & 9.2 & 14.26$\pm$0.01 & \\
& y2
& -13.3$\pm$0.1 & 16.9$\pm$0.3 & 2.1 & 16.8 & 14.91$\pm$0.02 & 15.03 \\
& y3 
& -37.3$\pm$3.2 & 41.3$\pm$4.3 & 17.7 & 37.3 & 13.55$\pm$0.06 & 14.53 \\ 
& 
& -21.0$\pm$0.1 & 10.7$\pm$0.2 & 5.4 & 9.2  & 14.26$\pm$0.01 &  \\ 
\hline
z & z1 
& 11.9$\pm$0.3 & 26.5$\pm$0.7 & 3.4 & 26.3 & 15.69$\pm$0.01 & 15.84 \\ 
& 
& 60.7$\pm$0.3 & 17.1$\pm$0.5 & 6.2 & 15.9 & 14.79$\pm$0.03 & \\
& z2 
& 31.6$\pm$1.8 & 47.6$\pm$1.8 & 16.7 & 44.6 & 14.23$\pm$0.05 & 15.54 \\ 
& 
& 32.8$\pm$0.3 & 13.3$\pm$0.8 & 3.4 & 12.9 & 15.39$\pm$0.06 & \\ 
& 
& 61.4$\pm$0.2 & 14.2$\pm$0.4 & 4.0 & 13.6 & 14.84$\pm$0.02 & \\
& z3 
& -7.7$\pm$2.2 & 17.8$\pm$4.6 & 16.6 & 6.4 & 12.83$\pm$0.10 & 15.21 \\ 
& 
& 42.2$\pm$0.2 & 25.1$\pm$0.4 & 3.7 & 24.8 & 15.06$\pm$0.20 & \\ 
& 
& 72.2$\pm$0.8 & 10.4$\pm$1.1 & 8.2 & 6.4 & 13.33$\pm$0.10 & \\
\hline
oblique & oblique1 
& -9.5$\pm$0.1 & 9.7$\pm$0.1 & 1.9 & 9.5 & 14.34$\pm$0.01 &  14.93 \\ 
& 
& 17.3$\pm$0.8 & 16.0$\pm$1.8 & 12.0 & 10.6 & 12.53$\pm$0.04 & \\
& oblique2 
& 5.6$\pm$0.1 & 17.1$\pm$0.3 & 2.1 & 17.0 & 15.13$\pm$0.02 &  15.12 \\ 
& 
& 42.3$\pm$0.4 & 10.4$\pm$1.0 & 4.0 & 9.6 & 13.20$\pm$0.03 & \\ 
& oblique3 
& 28.1$\pm$0.1 & 9.2$\pm$0.2 & 3.3 & 8.6 & 14.12$\pm$0.01 & 14.85 \\ 
& 
& 45.5$\pm$0.1 & 10.4$\pm$0.1 & 4.8 & 9.2 & 14.37$\pm$0.01 & \\  
&
& 59.3$\pm$1.1 & 58.9$\pm$1.5 & 27.6 & 52.0 & 13.82$\pm$0.01 & \\
\hline 
\multicolumn{8}{@{}l}{\parbox{0.9\textwidth}{\small\vspace{1mm}{Same as Table~\ref{tab:s1100}, but for Run 1 at 150 Myrs.}}} \\
\end{tabular} 
\label{tab:s1150} 

\end{table*}

\begin{table*}[!ht]
\caption{Run 1 at 200 Myrs} 
\hspace{-1.2cm}
\label{tab:my_complex_table_1kms_200Myr} 
\begin{tabular}{l l|*{5}{c}|c}
\hline
\multicolumn{2}{c|}{\multirow{2}{*}{Sight Lines}} & \multicolumn{5}{c|}{Results} & \multirow{2}{*}{$\log N(O~VI)_{sim}$} \\ 
\cline{3-7}
& & v$_{c}$ (km s$^{-1}$) & $b$ (km s$^{-1}$) & $b_T$ (km s$^{-1}$) & $b_{nt}$ (km s$^{-1}$) & $\log N(O~VI)$ \\ 
\hline
y & y1 
& -20.8$\pm$1.2 & 30.9$\pm$1.6 & 3.7 & 30.7 & 13.07$\pm$0.03 &  14.64 \\ 
&  
& -10.0$\pm$0.1 & 8.8$\pm$0.1 & 3.1 & 8.2 & 14.21$\pm$0.01 &  \\ 
& y2
& -25.6$\pm$1.9 & 18.8$\pm$2.0 & 2.3 & 18.7 & 14.50$\pm$0.07 &  15.06 \\ 
& 
& -6.1$\pm$0.7 & 10.7$\pm$1.4 & 2.2 & 10.5 & 14.83$\pm$0.17 & \\
& y3 
& -3.3$\pm$0.1 & 7.9$\pm$0.1 & 1.7 & 7.7 & 14.03$\pm$0.01 &  14.7 \\ 
\hline
z & z1 
& 46.4$\pm$0.3 & 36.1$\pm$1.5 & 2.4 & 36.0 & 16.60$\pm$0.18 &  16.47 \\
& z2 
& 70.2$\pm$0.1 & 25.4$\pm$0.4 & 3.5 & 25.2 & 15.95$\pm$0.01 &  15.83 \\ 
& z3 
& 96.3$\pm$0.2 & 31.6$\pm$0.4 & 18.3 & 25.8 & 13.59$\pm$0.01 &  13.47 \\
\hline
oblique & oblique1 
& -1.9$\pm$0.1 & 8.4$\pm$0.1 & 2.0 & 8.2 & 14.08$\pm$0.01 &  14.51 \\ 
& 
& 25.4$\pm$0.1 & 13.3$\pm$0.3 & 2.2 & 13.1 & 13.85$\pm$0.01 & \\
& oblique2 
& 11.8$\pm$0.1 & 7.5$\pm$0.2 & 2.2 & 7.2 & 14.11$\pm$0.02 &  14.78 \\ 
& 
& 21.0$\pm$0.7 & 19.3$\pm$0.7 & 17.9 & 7.2 & 13.88$\pm$0.03 & \\ 
& oblique3 
& 15.4$\pm$0.1 & 14.3$\pm$0.1 & 1.8 & 14.2 & 14.87$\pm$0.01 &  15.46 \\ 
& 
& 53.9$\pm$2.4 & 34.0$\pm$5.2 & 17.7 & 29.0 & 13.14$\pm$0.06 & \\
\hline 
\multicolumn{8}{@{}l}{\parbox{0.9\textwidth}{\small\vspace{1mm}{Same as Table~\ref{tab:s1100}, but for Run 1 at 200 Myrs.}}} \\
\end{tabular} 
\label{tab:s1200}
\end{table*}

\begin{figure*}[t]
    \centering
    \includegraphics[width=0.95\textwidth]{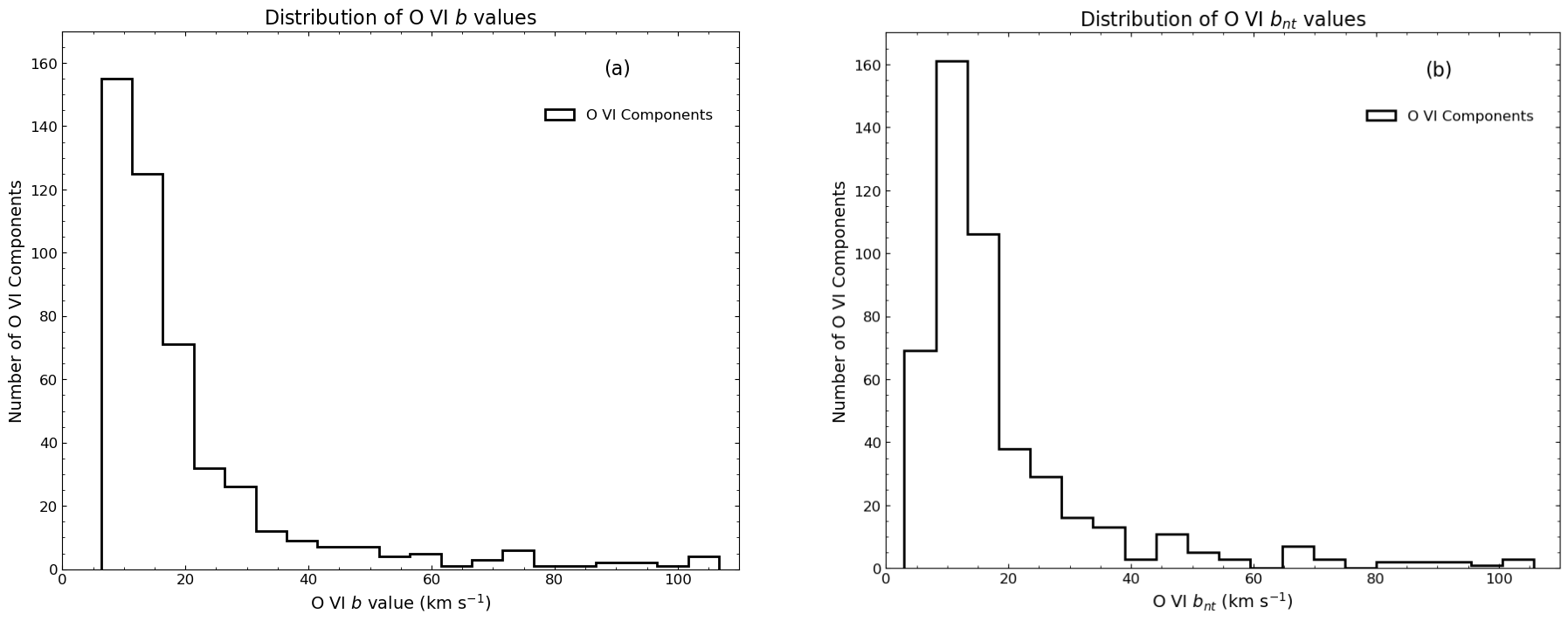}
    \caption{{Histograms of $b$ values (panel a) and $b_{nt}$ values (panel b) of {the \ion{O}{6} velocity components for all 270 simulated sight lines.}}}
    \label{fig:F15}
\end{figure*}

\begin{figure*}[t]
    \centering
    \includegraphics[width=0.95\textwidth]{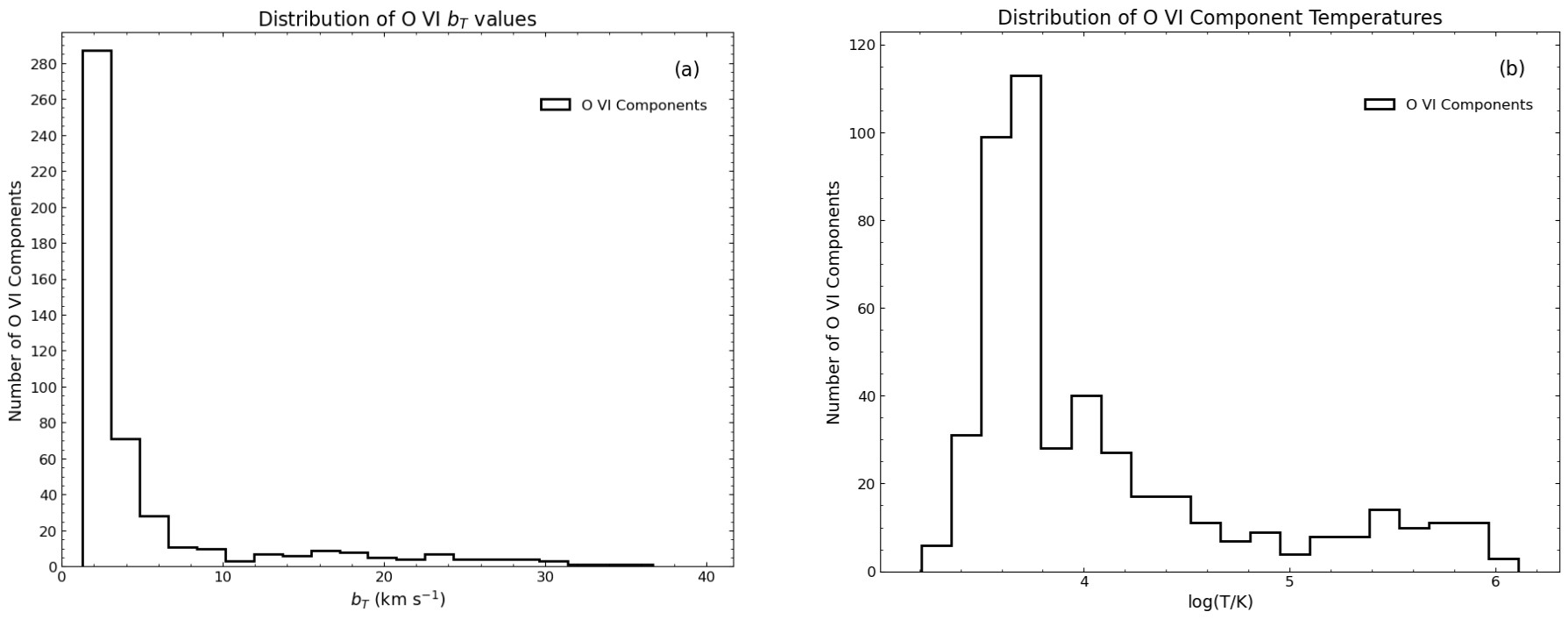}
    \caption{{Histograms of $b_{T}$ values (panel a) and temperatures (panel b) of the \ion{O}{6} velocity components for all 270 simulated sight lines.}}
    \label{fig:F16}
\end{figure*}

{The values of $b$ from Run 1 span from $\sim$6 km s$^{-1}$ to $\sim$93 km s$^{-1}$. The values of $b$ from all ten runs span from $\sim$6 km s$^{-1}$ to $\sim$107 km s$^{-1}$. {Figure~\ref{fig:F15} (a)} shows the histogram of our $b$ values from the 270 sight lines identified in Tables~\ref{tab:s1100}, \ref{tab:s1150}, \ref{tab:s1200} and Tables A1 to A27 for our 10 runs. Although our range is wide and the highest value suggests that non-thermal contributions are important, the figure shows that the majority of the $b$ values are smaller than 20 km s$^{-1}$. The median value, 14.2 km s$^{-1}$, also indicates that the narrow components dominate in our samples; very broad components are rare.}  
{\citet{Yang2025} also simulated fast moving clouds. Their plots of the \ion{O}{6} line width show a broad range of $b$ values, up to $\sim$100 km s$^{-1}$ and with a peak around 20 km s$^{-1}$. Our $b$ values also extend to such a value but our median is smaller and we have more cases of narrow \ion{O}{6} b values than they do owing to our use of NEI calculations.}

{Many sight lines have much wider or narrower line widths than expected from thermal broadening in CIE gas. In CIE \ion{O}{6} is most prevalent at T$_{CIE}$ = {2.9} $\times$ 10$^{5}$~K \citep{Gnat2007}, hence $b_{CIE}$=17.4~km~s$^{-1}$. For the sight lines that have $b$~$>>$~$b_{CIE}$, there is a clear trend that non-thermal broadening dominates the total broadening. In these regions, $b_{nt}$ is usually larger than $b_{T}$. Even for hot, mixed gas, the total line widths are still mostly due to non-thermal broadening from turbulent mixing and velocity gradients.  Even for the broad components, it is still rare for the temperature to reach an order of $\sim$10$^{6}$~K.} 
{The substantial non-thermal contributions are consistent with the existence of relative motions between the \ion{O}{6} clumps in Figure~\ref{fig:F3} and the velocity gradient in Figure~\ref{fig:F4}. }

{Sight lines that have $b$ $<<$ $b_{CIE}$ and a significant amount of \ion{O}{6} have gas temperatures below 10$^{5}$~K. 
In these sight lines, $b_{nt}$ is also greater than $b_{T}$, indicating that mixing and velocity gradients play more important roles than thermal broadening. 
Since many cells' temperatures along these sight lines are under 10$^{4}$~K, thermal broadening should have a very limited effect on $b$. The reason why \ion{O}{6} can exist in such cool places is that mixing brings highly ionized ambient material into contact with cool material. Heat is transferred from the hot gas to the cool gas. Since the density of the entrained ambient gas is much smaller than that of the cool cloud gas, the added heat does not greatly increase the temperature.} 

{Another important reason for these cells remaining cool is radiative cooling \citep{Kwak2010}. 
\citet{Kwak2010} analyzed the effects of non-equilibrium ionization and recombination in radiatively cooled, turbulent gas.} 
The gas temperatures and densities in their suite of simulations overlapped those of ours. {They found that radiative cooling is very effective and NEI increased the ionization fraction of \ion{O}{6} relative to the fraction in CIE gas at the same temperature.} 
In an example cell, whose temperature was $\sim$1.5$\times$10$^{5}$~K, the ionization fraction of \ion{O}{6} was $\sim$100 times greater in NEI calculations than in CIE calculations. On average, along examined sightlines, the NEI \ion{O}{6} ionization fraction was approximately twice the CIE \ion{O}{6} ionization fraction. 
We expect that these phenomena happen in our simulations as well. 
I.e., NEI, mixing and radiative cooling enable a significant amount of cool \ion{O}{6} to exist on some sight lines, resulting in $b_{T}$ $<$ $b_{nt}$.  

{Figure~\ref{fig:F15} (b) shows the histograms of our $b_{nt}$ values from the 270 sight lines identified in Tables~\ref{tab:s1100}, \ref{tab:s1150}, \ref{tab:s1200} and Tables A1 to A27 for our 10 runs. The range is from $\sim$4 km s$^{-1}$ to $\sim$106 km~s$^{-1}$. The median value of $b_{nt}$ is 13.5 km s$^{-1}$, indicating that is is rare to have extremely large values of $b_{nt}$.  By comparing the simulated $b$ and $b_{nt}$ histograms, it is clear that the non-thermal contribution accounts for most of the broadening effect in our simulations. } 

{Figure~\ref{fig:F16} shows the histograms of our $b_{T}$ and average temperature of the velocity components. The median value of $b_{T}$ is 2.5 km s$^{-1}$, which is much smaller than the value  associated with T$_{CIE}$ of 2.9$\times$10$^{5}$ K (i.e., 17.4 km s$^{-1}$). This result indicates that a substantial amount of \ion{O}{6} is out of CIE.  }


\section{Discussion} 
\label{sec:s4}

We investigated the impacts of NEI and mixing on the presence of \ion{O}{6} along simulated sight lines 
{and we found that sight lines passing through these cool \ion{O}{6}-rich regions tend to result in narrow \ion{O}{6} velocity components in the synthetic spectra. }
{Empirically, narrow line widths have been detected in observations toward HVCs near the Milky Way, as well as clouds in the CGM and IGM 
\citep{Sembach2003, Tripp2008, Qu2024, Savage2014}. Before discussing these observations, a comment about instrumentation is needed. The \citet{Tripp2008} study was mainly based on STIS data, whose high spectral resolution ($\sim$7~km~s$^{-1}$) is good for resolving relatively narrow absorbers. However, some of their sight lines were observed by FUSE, which has a spectral resolution of $\sim$15 km s$^{-1}$ \citep{Sahnow2000, Kaiser2009}. FUSE was also used for the \citet{Sembach2003} observations discussed below. The \citet{Savage2014} and \citet{Qu2024} studies discussed below used the Hubble Space Telescope Cosmic Origins Spectrograph (COS), which has a more modest resolution ($\sim$18~km~s$^{-1}$).} 

{The \citet{Tripp2008} STIS and FUSE study of low redshift intergalactic clouds produces excellent examples of observed \ion{O}{6} features that have narrow line widths. Their detailed analysis revealed a broad overall range of $b$ values from $\sim$4 to $\sim$76 km s$^{-1}$ (see their Table 7). Their study distinguished between two populations: intervening absorbers sampling the general IGM and proximate absorbers located near the background QSO. For the 70 robust intervening components, the median value of $b(\text{\ion{O}{6}})$ is 24~km~s$^{-1}$ and for the proximate absorbers the median value of $b(\text{\ion{O}{6}})$ is 17~km~s$^{-1}$.} 

{For a subset of intervening systems where \ion{O}{6} and \ion{H}{1} components were kinematically aligned, \citet{Tripp2008} determined the thermal and non-thermal contributions to $b$. They found that many \ion{O}{6} components are surprisingly cool, with 62\% of them implying temperatures of $T < 10^5$~K, and the non-thermal motions often dominate the line broadening. Their full sample yields a non-thermal broadening range of $b_{\text{NT}}$ from  6 to 75~km~s$^{-1}$. However, when the FUSE data were removed, due to its lower spectral resolution, from the \citet{Tripp2008} dataset, leaving only the STIS data, the $b_{\text{NT}}$ range shrank considerably to 10--33~km~s$^{-1}$ with a median value of 20.5 km s$^{-1}$ (see their Tables 2 and 7). The median value of $b_{NT}$ found by \citet{Tripp2008} is even larger than ours. Clearly, turbulence plays a very important role. }

{The \citet{Savage2014} COS study of CGM and IGM clouds also found many cool \ion{O}{6} clumps.} Their results revealed a broad range of $b(\text{\ion{O}{6}})$ values, spanning from $\sim$5 to $\sim$79 km s$^{-1}$, which is {approaching the range derived in our simulations ($\sim$6 to $\sim$107 km s$^{-1}$). Their median value of $b(\text{\ion{O}{6}})$ is 27 km s$^{-1}$. } They figured that these observed line widths ($b(\text{\ion{O}{6}}$)) were due to both a thermal component and a non-thermal component. {They were able to estimate the thermal contribution for 45 velocity components using a method along the lines of that of \citet{Tripp2008}. } \citet{Savage2014} found that 40 {of these} have temperatures $\le$ 10$^{5.5}$~K, the CIE temperature for \ion{O}{6}. Of those, 31 have temperatures $\le$ 10$^{4.8}$ K (see their Table 4). Although only 7 of their \ion{O}{6} features have $b(\text{\ion{O}{6}}) < b_{\text{CIE}}$, they found that the majority of their \ion{O}{6} dataset has temperatures which are below T$_{CIE}$. 

 {Table 4 in \citet{Savage2014} also listed the non-thermal contribution to the total $b$ values. For what they called photoionization absorbers, they got a range from 5 to 55 km s$^{-1}$ with a median value of 23 km s$^{-1}$. For what they called collisional ionization absorbers, they got a range from $<$10 to 56 km s$^{-1}$ with a median value of 29 km s$^{-1}$. These ranges differ from the one calculated from STIS data by \citet{Tripp2008}.}  

\citet{Qu2024} used COS to survey the warm-hot CGM around 
galaxies and galaxy groups. {They determined the \ion{O}{6} velocity dispersion ($\sigma_{OVI}$), where $b(\text{\ion{O}{6}})$ = $\sqrt{2}$ $\sigma_{OVI}$, for their observations. They identified 2 sight lines with $\sigma_{OVI}$} values that were smaller than that at the \ion{O}{6} CIE temperature ($\sim$12.3~km~s$^{-1}$). 
These studies mentioned that the observed broadening is a combination of both the thermal and non-thermal effects, but they did not explicitly calculate the non-thermal part. 

{\citet{Sembach2003}, who focused on \ion{O}{6} absorption features in Galactic high velocity gas, reported 2 sight lines with $b$ values smaller than $b_{\text{CIE}}$, although one of these is questionable. \citet{Sembach2003} also mentioned that FUSE has an instrumental broadening of $\sim$12 to $\sim$15~km~s$^{-1}$. This tends to make observed line widths greater than actual line widths. }

{Various authors have considered the ionization mechanism. \citet{Tripp2008} favored
photoionization from the extragalactic UV background, potentially in combination with some collisional ionization in hybrid models.} 
{
\citet{Oppenheimer2013} proposed that some intervening absorbers may be photoionized by nearby, but now inactive, AGN in fossil proximity zones. Such systems are expected to be cool ($T \sim 10^4$~K) and potentially less turbulent than gas processed by energetic shocks. 
} 


{\citet{Savage2014} classified the origins of the \ion{O}{6} absorption features as being photoionization, a combination of photoionization and collisional ionization, or just collisional ionization based on their deduced temperatures of the gas.} 
{They argued that some of the gas may be out of CIE, explaining some \ion{O}{6} in gas whose temperature is below T$_{CIE}$. The \ion{O}{6} may have yet to recombine to O~V and lower ions.
They noted that producing substantial column densities of \ion{O}{6} via photoionization from the extragalactic background requires the clouds to have very low densities and very long path lengths. 
}

{\citet{Sembach2003} concludes that photoionization is not likely to be the reason for the presence of \ion{O}{6} in the regions that they studied, i.e. HVCs near the Milky Way. The extragalactic background UV radiation is not strong enough to explain the quantity of observed \ion{O}{6} in those HVCs. In addition, their photoionization models would require very low cloud number densities, and thus very large sizes.}  

{Our simulations of mixing HVCs reproduce the narrow \ion{O}{6} line widths due to recombination in cooling gas without requiring photoionization. Furthermore, our simulation scenario does not require the cloud density to be extremely low (see Table~\ref{tab:params}). While HVCs may be similar to some physical structures in the Universe, they differ from others, for instance, shock heated low density (n$\sim$10$^{-5}$ cm$^{-3}$) intergalactic gas in the early Universe. \citet{Yoshikawa2006} modeled this gas, finding that most of the \ion{O}{6} ions are close to their CIE state because the time scales for \ion{O}{6} ionization and recombination in such low density gas are shorter than the cooling time scale and the Hubble time. The \ion{O}{7} and \ion{O}{8} recombine much slower, so do not readily replenish the \ion{O}{6} population. 
\citet{Yoshikawa2006} cosmological simulations yield similar results as their simple shock heated model, except near cosmological filaments, where the \ion{O}{6} ionization fraction was greater than that of CIE. }

{In confirmation of oxygen being drastically out of equilibrium in our simulations, we find a notable trend in the oxygen ion population in cells with temperatures below 10$^{4}$ K. Most of these cells have more \ion{O}{7} ions than \ion{O}{6} ions. The \ion{O}{7} ions must have originated in the hot, highly ionized ambient gas. That gas has cooled, but much of the \ion{O}{7} has yet to recombine. The cool \ion{O}{6} in those cells is likely the result of recombination from such higher ionization states.} 


{To explain this process, consider hot diffuse ambient gas, which has a temperature of 2 $\times$ 10$^{6}$~K and is rich in \ion{O}{7} and \ion{O}{8}. It mixes with cold dense cloud gas which lowers its temperature, and it radiatively cools, causing its temperature to drop further.
The oxygen ion population does not immediately reach equilibrium, due to NEI effects. Instead, there is a temporal delay before the ionization and recombination rates approach their expected values. This process results in a significant amount of \ion{O}{6} that was produced through recombination from \ion{O}{7}, but has not yet recombined to \ion{O}{5}. This explains why there is an overabundance of \ion{O}{6} in low temperature cells as seen in our simulations. }

\section{Conclusion}
\label{sec:s5}

We develop a method to calculate the Doppler broadening parameter, $b$, in high velocity clouds that were simulated by the FLASH code. In our simulations, cool clouds quickly move through and mix with hot ambient gas {without photoionization}. Our results reveal plentiful amounts of \ion{O}{6} and a wide range of $b$ values{, most of which are narrow.} {On most sight lines, the non-thermal contribution is larger than the thermal contribution to $b$.} 

The narrow b values are of particular interest. Here we provide an explanation for them. {Mixing with cool cloud gas and then radiative cooling rapidly lowers the temperature of the ambient gas.} However, due to NEI and mixing, the ionization and recombination rates do not immediately reach equilibrium at the new temperature. This temporal delay results in recombination rates from \ion{O}{7} to \ion{O}{6} and from \ion{O}{6} to O V that are lower than those under CIE conditions. {Consequently in relatively cool regions}, an overabundance of \ion{O}{6} is observed as it remains in a transitional state, waiting for recombination to O V. These processes may explain the narrow line widths observed in various astrophysical environments, such as HVCs, CGM, and IGM.

\section{Acknowledgment} 
\label{sec:s6} 
We greatly appreciate the referee whose helpful suggestions broadened the scope of this paper and improved its readability.
We thank Shan-ho Tsai of the Georgia Advanced Computing Resource Center (GACRC) for her assistance. These simulations were performed on the GACRC computer clusters.

\appendix

\setcounter{table}{0} 
\renewcommand{\thetable}{A.\arabic{table}} 

\begin{table*}[!ht]
\caption{Run 2 at 100 Myrs}
\hspace{-1.2cm}
\label{tab:my_complex_table_run2_100Myr}
\begin{tabular}{l l|*{5}{c}|c}
\hline
\multicolumn{2}{c|}{\multirow{2}{*}{Sight Lines}} & \multicolumn{5}{c|}{Results} & \multirow{2}{*}{$\log N(O~VI)_{sim}$} \\ 
\cline{3-7}
& & v$_{c}$ (km s$^{-1}$) & $b$ (km s$^{-1}$) & $b_T$ (km s$^{-1}$) & $b_{nt}$ (km s$^{-1}$) & $\log N(O~VI)$ \\ 
\hline
y & y1 
& -0.7$\pm$0.1 & 7.8$\pm$0.1 & 1.9  & 7.6 & 14.06$\pm$0.01 &  15.08 \\ 
& y2 
& -3.7$\pm$0.1 & 11.7$\pm$0.1 & 1.9  & 11.5 & 14.58$\pm$0.01  & 15.16 \\ 
& y3  
& -11.5$\pm$0.2 & 30.7$\pm$0.4 &  18.7 & 24.3 & 13.11$\pm$0.01 & 13.91 \\ 
&   
& -1.9$\pm$0.1 & 7.1$\pm$0.1 & 5.7  & 4.2 & 13.69$\pm$0.01 &  \\ 

\hline
z & 
z1  
& -17.4$\pm$8.7 & 54.4$\pm$11.3 &  8.7 & 53.7 & 13.07$\pm$0.13 & 15.87 \\ 
&   
& -1.4$\pm$0.1 & 15.5$\pm$0.2 &  2.3  & 15.3 & 15.03$\pm$0.01 &  \\
& z2  
& -1.5$\pm$0.1 & 13.8$\pm$0.1 & 2.1  & 13.6 & 14.65$\pm$0.01 & 15.28 \\
& z3  
&  -5.4$\pm$5.1 & 17.6$\pm$2.5 & 2.5  & 17.4 & 14.24$\pm$0.32 &  15.19 \\ 
& 
& 1.8$\pm$0.2 & 11.3$\pm$1.7 & 2.4 & 11.0 & 14.76$\pm$0.04 & \\
\hline
oblique & oblique1  
& -8.7$\pm$0.1 & 8.6$\pm$0.1 & 1.8  & 8.4 & 14.15$\pm$0.01  & 15.02  \\ 
& oblique2  
& 1.1$\pm$0.1 & 14.8$\pm$0.1 &  1.9  & 14.7 & 14.88$\pm$0.01  &  15.40 \\ 
& oblique3  
& 0.6$\pm$0.1 & 7.9$\pm$0.1 & 2.1  & 7.6 & 13.98$\pm$0.01 & 14.93 \\
\hline 
\multicolumn{8}{@{}l}{\parbox{0.9\textwidth}{\small\vspace{1mm}{Same as Table~\ref{tab:s1100}, but for Run 2 at 100 Myrs.}}} \\
\end{tabular}
\end{table*} 

\begin{table*}[!ht]
\caption{Run 2 at 150 Myrs}
\hspace{-1.2cm}
\label{tab:my_complex_table_run2_150Myrs}
\begin{tabular}{l l|*{5}{c}|c}
\hline
\multicolumn{2}{c|}{\multirow{2}{*}{Sight Lines}} & \multicolumn{5}{c|}{Results} & \multirow{2}{*}{$\log N(O~VI)_{sim}$} \\ 
\cline{3-7}
& & v$_{c}$ (km s$^{-1}$) & $b$ (km s$^{-1}$) & $b_T$ (km s$^{-1}$) & $b_{nt}$ (km s$^{-1}$) & $\log N(O~VI)$ \\ 
\hline
y & y1 
& -0.2$\pm$0.1 & 10.3$\pm$0.1 &  2.1 & 10.1 & 14.38$\pm$0.01 &  15.16 \\
& y2 
& -4.8$\pm$0.1 & 11.7$\pm$0.1 & 1.8  & 11.6 & 14.57$\pm$0.01 &  15.13 \\ 
& y3  
& -8.1$\pm$0.1 & 13.6$\pm$0.2 &  2.2 & 13.4 & 14.85$\pm$0.02 &  15.47 \\ 
\hline
z & 
z1  
& 6.4$\pm$0.1 & 18.4$\pm$0.5 & 2.4  & 18.2 & 15.59$\pm$0.04 &  16.18 \\ 
&   
& 19.0$\pm$1.0 & 33.9$\pm$4.0 &  3.0  & 33.8 & 13.79$\pm$0.32 &  \\
& z2  
& 7.2$\pm$0.1 & 16.6$\pm$0.2 & 2.2  & 16.5 & 15.39$\pm$0.02 &  15.72 \\ 
& z3  
&  -13.6$\pm$9.9 & 36.3$\pm$13.9 & 22.6  & 28.4 & 13.02$\pm$0.20 & 15.51 \\ 
& 
& 13.5$\pm$0.1 & 15.0$\pm$0.1 & 2.5 & 14.8 & 14.87$\pm$0.01 & \\
\hline
oblique & oblique1  
& 1.1$\pm$0.1 & 12.1$\pm$0.3 & 2.0  & 11.9 & 14.59$\pm$0.02 &  15.00 \\ 
&  
& 26.6$\pm$1.1 & 9.4$\pm$1.6 &  3.0 & 8.9 & 12.95$\pm$0.08 & \\
& oblique2  
& -0.6$\pm$0.1 & 9.5$\pm$0.1 &  1.9  & 9.3 & 14.32$\pm$0.01  & 15.31 \\ 
& oblique3  
& 7.5$\pm$0.2 & 12.2$\pm$0.8 &  2.6 & 11.9 & 14.77$\pm$0.06 &  15.57 \\ 
&  
& 42.0$\pm$0.2 & 17.3$\pm$0.5 &  3.4  & 17.0 & 14.86$\pm$0.03 & \\
\hline 
\multicolumn{8}{@{}l}{\parbox{0.9\textwidth}{\small\vspace{1mm}{Same as Table~\ref{tab:s1100}, but for Run 2 at 150 Myrs.}}} \\
\end{tabular}
\end{table*}  

\begin{table*}[!ht]
\caption{Run 2 at 200 Myrs}
\hspace{-1.2cm}
\label{tab:my_complex_table_run2_200Myrs}
\begin{tabular}{l l|*{5}{c}|c}
\hline
\multicolumn{2}{c|}{\multirow{2}{*}{Sight Lines}} & \multicolumn{5}{c|}{Results} & \multirow{2}{*}{$\log N(O~VI)_{sim}$} \\ 
\cline{3-7}
& & v$_{c}$ (km s$^{-1}$) & $b$ (km s$^{-1}$) & $b_T$ (km s$^{-1}$) & $b_{nt}$ (km s$^{-1}$) & $\log N(O~VI)$ \\ 
\hline
y & y1 
& 4.3$\pm$0.1 & 10.6$\pm$0.1 & 2.0  & 10.4 & 14.52$\pm$0.01 &  15.37 \\ 
& y2 
& -6.3$\pm$0.1 & 10.4$\pm$0.1 &  2.1 & 10.2 & 14.36$\pm$0.01 &  15.35 \\ 
& y3  
& 2.1$\pm$0.1 & 8.1$\pm$0.1 & 1.5  & 8.0 & 14.12$\pm$0.01 & 15.10 \\ 
\hline
z & 
z1  
& 15.0$\pm$0.1 & 19.4$\pm$0.2 &  2.2 & 19.3 & 15.35$\pm$0.02 & 16.39 \\
& z2  
& 14.8$\pm$0.1 & 18.3$\pm$0.1 & 2.1  & 18.2 & 15.32$\pm$0.01 &  16.21 \\
& z3  
& 5.9$\pm$2.3  & 37.1$\pm$2.6 & 36.7  & 5.4 & 13.28$\pm$0.05 &  15.55 \\ 
& 
& 25.5$\pm$0.1 & 13.5$\pm$0.1 & 2.6 & 13.2 & 14.66$\pm$0.01 & \\
\hline
oblique & oblique1  
& -3.5$\pm$0.1 & 10.4$\pm$0.1 &  2.0 & 10.2 & 14.46$\pm$0.01 & 15.32 \\ 
& oblique2  
& 10.3$\pm$0.1 & 13.4$\pm$0.1 &  2.1  & 13.2 & 14.69$\pm$0.01  & 15.36 \\ 
& oblique3  
& 4.4$\pm$0.1 & 9.8$\pm$0.1 & 1.5  & 9.7 & 14.34$\pm$0.01 & 15.17 \\
\hline 
\multicolumn{8}{@{}l}{\parbox{0.9\textwidth}{\small\vspace{1mm}{Same as Table~\ref{tab:s1100}, but for Run 2 at 200 Myrs.}}} \\
\end{tabular}
\end{table*}  

\begin{table*}[!ht]
\caption{Run 3 at 100 Myrs}
\hspace{-1.2cm}
\label{tab:my_complex_table_run3_100Myr}
\begin{tabular}{l l|*{5}{c}|c}
\hline
\multicolumn{2}{c|}{\multirow{2}{*}{Sight Lines}} & \multicolumn{5}{c|}{Results} & \multirow{2}{*}{$\log N(O~VI)_{sim}$} \\ 
\cline{3-7}
& & v$_{c}$ (km s$^{-1}$) & $b$ (km s$^{-1}$) & $b_T$ (km s$^{-1}$) & $b_{nt}$ (km s$^{-1}$) & $\log N(O~VI)$ \\ 
\hline
y & y1 
& -5.1$\pm$0.3 & 18.5$\pm$0.6 &  6.4 & 17.4 & 13.44$\pm$0.03 & 14.70 \\ 
&   
& -3.6$\pm$0.1 & 6.8$\pm$0.2 & 2.4  & 6.4 & 14.03$\pm$0.01 &  \\ 
& y2 
& -17.6$\pm$2.1 & 23.7$\pm$2.5 & 2.5  & 23.6 & 14.08$\pm$0.06 & 14.93 \\ 
& 
& -0.7$\pm$0.2 & 10.7$\pm$0.4 & 2.0 & 10.5 & 14.41$\pm$0.03 & \\
& y3  
& -21.9$\pm$1.3 & 36.7$\pm$1.9 & 26.4  & 25.5 & 13.19$\pm$0.03 & 15.14 \\ 
&   
& -2.6$\pm$0.1 & 8.9$\pm$0.1 & 2.2  & 8.6 & 14.22$\pm$0.01 &  \\
\hline
z & 
z1  
& 27.8$\pm$0.1 & 26.6$\pm$0.5 & 2.4  & 26.5 & 16.36$\pm$0.15 &  16.21 \\
& z2  
& 32.4$\pm$0.2 & 22.2$\pm$0.7 & 3.6  & 21.9 & 15.27$\pm$0.01 &  15.32 \\ 
&  
& 77.1$\pm$2.2 & 18.1$\pm$4.6 &  15.3 & 9.7 & 13.34$\pm$0.09 & \\
& z3  
&  26.5$\pm$0.7 & 10.1$\pm$0.7 & 3.0  & 9.6 & 14.21$\pm$0.06 &  14.71 \\ 
& 
& 45.6$\pm$0.6 & 16.4$\pm$0.8 & 3.8 & 16.0 & 14.78$\pm$0.03 & \\
\hline
oblique & oblique1  
& -2.6$\pm$0.1 & 12.7$\pm$0.2 & 2.2  & 12.5 & 14.46$\pm$0.01 & 14.99 \\
& oblique2  
& 4.0$\pm$0.1 & 13.2$\pm$0.1 &  1.9  & 13.1 & 14.63$\pm$0.01  & 14.97 \\ 
& oblique3  
& 14.4$\pm$0.1 & 11.6$\pm$0.1 &  2.2 & 11.4 & 14.48$\pm$0.01 &  15.25 \\ 
&  
& 32.5$\pm$0.9 & 13.5$\pm$1.4 & 2.4   & 13.3 & 12.68$\pm$0.05 & \\
\hline 
\multicolumn{8}{@{}l}{\parbox{0.9\textwidth}{\small\vspace{1mm}{Same as Table~\ref{tab:s1100}, but for Run 3 at 100 Myrs.}}} \\
\end{tabular}
\end{table*}  

\begin{table*}[!ht]
\caption{Run 3 at 150 Myrs}
\hspace{-1.2cm}
\label{tab:my_complex_table_run3_150Myr}
\begin{tabular}{l l|*{5}{c}|c}
\hline
\multicolumn{2}{c|}{\multirow{2}{*}{Sight Lines}} & \multicolumn{5}{c|}{Results} & \multirow{2}{*}{$\log N(O~VI)_{sim}$} \\ 
\cline{3-7}
& & v$_{c}$ (km s$^{-1}$) & $b$ (km s$^{-1}$) & $b_T$ (km s$^{-1}$) & $b_{nt}$ (km s$^{-1}$) & $\log N(O~VI)$ \\ 
\hline
y & y1 
& -6.5$\pm$0.2 & 10.5$\pm$0.4 &  2.2 & 10.3  & 14.22$\pm$0.02 & 14.34 \\ 
& y2 
& -11.6$\pm$0.6 & 15.3$\pm$0.5 & 2.3  & 15.1 & 13.81$\pm$0.03 & 14.71 \\ 
& 
& -2.7$\pm$0.1 & 6.7$\pm$0.2 & 2.2 & 6.3 & 13.95$\pm$0.02 & \\
& y3  
& -10.9$\pm$0.2 & 24.4$\pm$0.3 & 9.0  & 22.7 & 12.94$\pm$0.01 & 15.05 \\ 
&   
& -2.7$\pm$0.1 & 8.2$\pm$0.1 &  2.1 & 7.9 & 14.11$\pm$0.01 &  \\
\hline
z & 
z1  
& 49.5$\pm$0.1 & 24.6$\pm$0.2 &  2.4 & 24.5 & 16.08$\pm$0.03 &  16.18 \\ 
& z2  
& 51.0$\pm$0.3 & 19.1$\pm$1.0 & 3.2  & 18.8 & 15.06$\pm$0.05 &  15.66 \\ 
&  
& 51.1$\pm$1.2 & 48.2$\pm$4.7 &  14.1 & 46.1 & 14.31$\pm$0.08 & \\
& z3  
&  49.0$\pm$0.3 & 10.7$\pm$0.4 &  5.2 & 9.4 & 14.31$\pm$0.05 &  15.71 \\ 
& 
& 57.9$\pm$2.3 & 50.0$\pm$7.8 & 15.2 & 47.6 & 13.54$\pm$0.12 & \\ 
& 
& 61.8$\pm$1.8 & 19.5$\pm$2.0 & 5.6 & 18.7 & 14.15$\pm$0.09 & \\
\hline
oblique & oblique1  
& 7.0$\pm$0.1 & 11.9$\pm$0.1 &  2.2 & 11.7 & 14.51$\pm$0.01 &  15.00 \\ 
&  
& 28.9$\pm$0.1 & 8.3$\pm$0.3 &  3.1 & 7.7 & 13.26$\pm$0.01 & \\
& oblique2  
& 16.8$\pm$0.1 & 14.8$\pm$0.2 &  2.3  & 14.6 & 14.62$\pm$0.01  & 14.72 \\ 
& 
& 53.5$\pm$0.9 & 12.0$\pm$1.7 & 10.4  & 6.0 & 13.13$\pm$0.06 & \\ 
& oblique3  
& 15.6$\pm$0.1 & 8.7$\pm$0.1 & 1.8  & 8.5 & 14.21$\pm$0.01 &  14.97 \\
\hline 
\multicolumn{8}{@{}l}{\parbox{0.9\textwidth}{\small\vspace{1mm}{Same as Table~\ref{tab:s1100}, but for Run 3 at 150 Myrs.}}} \\
\end{tabular}
\end{table*}  

\begin{table*}[!ht]
\caption{Run 3 at 200 Myrs}
\hspace{-1.2cm}
\label{tab:my_complex_table_run3_200Myr}
\begin{tabular}{l l|*{5}{c}|c}
\hline
\multicolumn{2}{c|}{\multirow{2}{*}{Sight Lines}} & \multicolumn{5}{c|}{Results} & \multirow{2}{*}{$\log N(O~VI)_{sim}$} \\ 
\cline{3-7}
& & v$_{c}$ (km s$^{-1}$) & $b$ (km s$^{-1}$) & $b_T$ (km s$^{-1}$) & $b_{nt}$ (km s$^{-1}$) & $\log N(O~VI)$ \\ 
\hline
y & y1 
& -5.2$\pm$0.2 & 11.4$\pm$0.2 &  2.0 & 11.2 & 14.33$\pm$0.01 &  14.78 \\
& y2 
& 1.1$\pm$0.1 & 10.0$\pm$0.2 &  1.9 & 9.8 & 14.36$\pm$0.01 & 14.96 \\ 
& y3  
& -0.9$\pm$0.1 & 9.5$\pm$0.1 & 2.2  & 9.2 & 14.33$\pm$0.01 &  14.88 \\  
\hline
z & 
z1  
& 66.0$\pm$0.1 & 19.1$\pm$0.2 & 2.3  & 19.0 & 15.65$\pm$0.03 &  16.14 \\ 
& z2  
& 69.7$\pm$0.1 & 18.6$\pm$0.2 &  2.5 & 18.4 & 15.33$\pm$0.02 &  15.98 \\
& z3  
& 69.2$\pm$0.1  & 17.5$\pm$0.2 &  2.2 & 17.4 & 15.44$\pm$0.03 & 15.96 \\
\hline
oblique & oblique1  
& 19.7$\pm$0.1 & 13.8$\pm$0.3 & 1.9  & 13.7 & 14.63$\pm$0.02 & 14.77 \\ 
&  
& 44.6$\pm$0.6 & 7.3$\pm$0.9 & 2.8  & 6.7 & 13.28$\pm$0.04 & \\
& oblique2  
& 22.6$\pm$0.1 & 11.5$\pm$0.1 &  1.9  & 11.3 &  14.47$\pm$0.01 & 14.78 \\ 
& 
& 37.7$\pm$0.1 & 7.1$\pm$0.1 &  2.7 & 6.6 & 13.47$\pm$0.01 & \\ 
& oblique3  
& 26.1$\pm$0.1 & 8.6$\pm$0.1 &  2.1 & 8.3 & 14.15$\pm$0.01 & 14.78 \\
\hline 
\multicolumn{8}{@{}l}{\parbox{0.9\textwidth}{\small\vspace{1mm}{Same as Table~\ref{tab:s1100}, but for Run 3 at 200 Myrs.}}} \\
\end{tabular}
\end{table*}

\begin{table*}[!ht]
\caption{Run 4 at 100 Myrs}
\hspace{-1.2cm}
\label{tab:my_complex_table_run4_100Myr}
\begin{tabular}{l l|*{5}{c}|c}
\hline
\multicolumn{2}{c|}{\multirow{2}{*}{Sight Lines}} & \multicolumn{5}{c|}{Results} & \multirow{2}{*}{$\log N(O~VI)_{sim}$} \\ 
\cline{3-7}
& & v$_{c}$ (km s$^{-1}$) & $b$ (km s$^{-1}$) & $b_T$ (km s$^{-1}$) & $b_{nt}$ (km s$^{-1}$) & $\log N(O~VI)$ \\ 
\hline
y & y1 
& -10.8$\pm$1.8 & 17.7$\pm$1.3 &  16.5 & 6.4 & 13.38$\pm$0.08 & 15.13 \\ 
&   
& -1.5$\pm$0.1 & 8.9$\pm$0.1 &  2.7 & 8.5 & 14.18$\pm$0.01 &  \\ 
& y2 
& -45.5$\pm$0.4 & 13.7$\pm$0.7 & 2.5  & 13.5 & 14.15$\pm$0.02 & 15.02 \\ 
& 
& -5.4$\pm$0.2 & 8.3$\pm$0.4 & 2.1 & 8.0 & 14.10$\pm$0.03 & \\
& y3  
& -21.7$\pm$0.1 & 8.4$\pm$0.2 &  5.6 & 6.3 & 13.51$\pm$0.01 & 14.74 \\ 
&   
& -19.8$\pm$0.5 & 32.4$\pm$0.9 & 14.5  & 29.0 & 13.53$\pm$0.02 &  \\ 
&   
& -6.1$\pm$0.1 & 8.1$\pm$0.1 &  3.8 & 7.2 & 14.18$\pm$0.01 &  \\ 
\hline
z & 
z1  
& 36.3$\pm$0.4 & 15.3$\pm$1.1 & 2.5  & 15.1 & 14.94$\pm$0.09 & 15.92 \\ 
&   
& 93.8$\pm$0.8 & 28.2$\pm$1.7 &  6.2  & 27.5 & 15.27$\pm$0.13 &  \\ 
&   
& 133.1$\pm$5.7 & 106.7$\pm$5.8 &  14.5 & 105.7 & 14.92$\pm$0.04 &  \\ 
& z2  
& 36.7$\pm$2.9 & 14.9$\pm$2.2 &  2.2 & 14.7 & 14.97$\pm$0.23 & 15.79 \\ 
&
& 75.0$\pm$2.1 & 42.4$\pm$1.7 & 2.3 & 42.3 & 15.32$\pm$0.04 & \\
& z3  
& 138.4$\pm$0.5  & 11.8$\pm$1.2 & 11.4  & 3.0 & 14.16$\pm$0.06 & 14.48  \\ 
& 
& 152.2$\pm$4.0 & 30.2$\pm$4.0 & 26.0 & 15.4 & 13.96$\pm$0.10 & \\
\hline
oblique & oblique1  
& 26.3$\pm$0.1 & 17.1$\pm$0.2 & 2.1  & 17.0 & 15.13$\pm$0.01 &  15.40 \\ 
&  
& 55.6$\pm$1.7 & 23.2$\pm$2.4 & 13.1  & 19.1 & 13.55$\pm$0.05 & \\
& oblique2  
& 26.3$\pm$0.1 & 11.2$\pm$0.2 & 2.4   & 10.9 & 14.40$\pm$0.01  & 15.23 \\ 
& 
& 81.3$\pm$0.3 & 13.0$\pm$0.6 & 9.3  & 9.1 & 13.78$\pm$0.02 & \\ 
&  
& 111.4$\pm$13.6 & 104.7$\pm$8.0 & 26.6  & 101.3 & 13.92$\pm$0.03 & \\
& oblique3  
& 37.4$\pm$0.2 & 9.0$\pm$1.3 &  2.1 & 8.8 & 14.28$\pm$0.03 &  14.76 \\ 
&  
& 54.3$\pm$1.0 & 9.3$\pm$3.1 &  5.5  & 7.5 & 13.55$\pm$0.09 &  \\ 
& 
& 76.4$\pm$4.1 & 62.2$\pm$5.8 & 29.7 & 54.7 & 13.89$\pm$0.05 & \\
\hline 
\multicolumn{8}{@{}l}{\parbox{0.9\textwidth}{\small\vspace{1mm}{Same as Table~\ref{tab:s1100}, but for Run 4 at 100 Myrs.}}} \\
\end{tabular}
\end{table*}

\begin{table*}[!ht]
\caption{Run 4 at 150 Myrs}
\hspace{-1.2cm}
\label{tab:my_complex_table_run4_150Myr}
\begin{tabular}{l l|*{5}{c}|c}
\hline
\multicolumn{2}{c|}{\multirow{2}{*}{Sight Lines}} & \multicolumn{5}{c|}{Results} & \multirow{2}{*}{$\log N(O~VI)_{sim}$} \\ 
\cline{3-7}
& & v$_{c}$ (km s$^{-1}$) & $b$ (km s$^{-1}$) & $b_T$ (km s$^{-1}$) & $b_{nt}$ (km s$^{-1}$) & $\log N(O~VI)$ \\ 
\hline
y & y1 
& -2.0$\pm$0.1 & 15.1$\pm$0.4 &  2.1 & 15.0 & 14.79$\pm$0.03 &  15.41 \\
& y2 
& -30.2$\pm$0.4 & 8.7$\pm$0.6 & 3.6  & 7.9 & 13.27$\pm$0.03 &  15.11 \\ 
& 
& -1.1$\pm$0.1 & 12.7$\pm$0.2 & 2.1 & 12.5 & 14.48$\pm$0.01 & \\
& y3  
& -14.2$\pm$1.8 & 23.6$\pm$2.2 &  15.4 & 17.9 & 13.20$\pm$0.06 & 15.26  \\ 
&   
& 2.7$\pm$0.1 & 10.8$\pm$0.1 & 2.5  & 10.5 & 14.41$\pm$0.01 &  \\
\hline
z & 
z1  
& 61.3$\pm$11.1 & 20.8$\pm$5.4 & 2.1  & 20.7 & 16.10$\pm$0.86 & 16.31 \\ 
&   
& 120.0$\pm$5.1 & 67.3$\pm$3.6 &  3.2  & 67.2 & 15.63$\pm$0.07 &  \\
& z2  
& 93.5$\pm$0.3 & 47.5$\pm$1.5 & 3.3  & 47.4 & 16.07$\pm$0.09 &  16.12 \\ 
&  
& 208.6$\pm$2.6 & 49.3$\pm$5.5 & 18.4  & 45.7 & 14.23$\pm$0.04 & \\
& z3  
& 69.5$\pm$2.9  & 21.6$\pm$3.2 &  2.4 & 21.5 & 14.79$\pm$0.15 & 14.94 \\ 
& 
& 96.7$\pm$8.8 & 35.6$\pm$8.1 & 4.4 & 35.3 & 14.66$\pm$0.20 & \\
& 
& 221.6$\pm$2.5 & 39.4$\pm$4.9 & 17.4 & 35.3 & 14.05$\pm$0.05 & \\
\hline
oblique & oblique1  
& 25.0$\pm$0.2 & 13.5$\pm$0.3 &  2.1 & 13.3 & 14.69$\pm$0.02 &  15.01 \\ 
&  
& 54.8$\pm$1.5 & 26.7$\pm$2.7 & 5.6  & 26.1 & 13.84$\pm$0.04 & \\
& oblique2  
& 34.5$\pm$0.1 & 12.4$\pm$0.1 &  2.2  & 12.2 &  14.61$\pm$0.01 & 15.21 \\ 
& oblique3  
& 58.3$\pm$0.1 & 10.8$\pm$0.3 & 2.2  & 10.6 & 14.42$\pm$0.01 & 15.15 \\ 
&  
& 65.7$\pm$1.5 & 30.0$\pm$2.3 & 6.1   & 29.4 & 13.65$\pm$0.06 & \\
\hline 
\multicolumn{8}{@{}l}{\parbox{0.9\textwidth}{\small\vspace{1mm}{Same as Table~\ref{tab:s1100}, but for Run 4 at 150 Myrs.}}} \\
\end{tabular}
\end{table*}

\begin{table*}[!ht]
\caption{Run 4 at 200 Myrs}
\hspace{-1.2cm}
\label{tab:my_complex_table_run4_200Myr}
\begin{tabular}{l l|*{5}{c}|c}
\hline
\multicolumn{2}{c|}{\multirow{2}{*}{Sight Lines}} & \multicolumn{5}{c|}{Results} & \multirow{2}{*}{$\log N(O~VI)_{sim}$} \\ 
\cline{3-7}
& & v$_{c}$ (km s$^{-1}$) & $b$ (km s$^{-1}$) & $b_T$ (km s$^{-1}$) & $b_{nt}$ (km s$^{-1}$) & $\log N(O~VI)$ \\ 
\hline
y & y1 
& 4.5$\pm$0.1 & 12.4$\pm$0.1 &  2.2 & 12.2 & 14.60$\pm$0.01 & 15.30 \\
& y2 
& -1.2$\pm$0.1 & 9.5$\pm$0.1 &  1.7 & 9.3 & 14.33$\pm$0.01 &  14.93 \\ 
& y3  
& -31.4$\pm$2.7 & 23.5$\pm$5.1 & 16.3  & 16.9 & 13.86$\pm$0.09 & 15.01 \\ 
&   
& -5.5$\pm$0.3 & 12.0$\pm$1.1 &  2.3 & 11.8 & 14.71$\pm$0.09 &  \\
\hline
z & 
z1  
& 58.6$\pm$0.8 & 13.9$\pm$1.0 & 2.2  & 13.7 & 15.76$\pm$0.18 &  16.31 \\ 
&   
& 110.7$\pm$0.9 & 27.0$\pm$1.2 & 3.3   & 26.8 & 16.16$\pm$0.02 &  \\ 
&   
& 171.3$\pm$2.4 & 57.6$\pm$3.3 &  24.8 & 52.0 & 14.27$\pm$0.03 &  \\ 
& z2  
& 65.1$\pm$0.2 & 24.8$\pm$1.0 & 2.3  & 24.7 & 15.56$\pm$0.08 &  15.81 \\ 
&  
& 130.8$\pm$0.4 & 23.9$\pm$1.3 & 8.7  & 22.3 & 14.64$\pm$0.03 & \\ 
& 
& 170.3$\pm$9.9 & 75.4$\pm$12.3 & 23.5 & 71.6 & 14.30$\pm$0.10 & \\ 
& z3  
&  101.3$\pm$0.3 & 31.1$\pm$1.1 & 3.3  & 30.9 & 15.70$\pm$0.01 & 15.58 \\
\hline
oblique & oblique1  
& 16.4$\pm$0.2 & 10.3$\pm$0.4 & 1.4  & 10.2 & 14.28$\pm$0.02 &  14.84 \\ 
&  
& 41.5$\pm$0.4 & 15.9$\pm$0.9 & 1.7  & 15.8 & 13.94$\pm$0.02 & \\ 
& 
& 74.9$\pm$1.0 & 8.5$\pm$1.5 & 3.6 & 7.7 & 13.09$\pm$0.07 & \\
& oblique2  
& 37.1$\pm$0.3 & 11.9$\pm$0.4 &  3.3  & 11.4 &  14.51$\pm$0.03 & 15.12 \\ 
& 
&  66.3$\pm$0.7 & 26.1$\pm$1.2 & 4.6  & 25.7 & 14.36$\pm$0.02 & \\
& oblique3  
& 56.2$\pm$0.6 & 9.4$\pm$1.1 &  2.2 & 9.1 & 14.29$\pm$0.08 &  15.24 \\ 
&  
& 72.5$\pm$1.9 & 23.6$\pm$2.2  &  2.6  & 23.5 & 14.42$\pm$0.06 & \\
\hline 
\multicolumn{8}{@{}l}{\parbox{0.9\textwidth}{\small\vspace{1mm}{Same as Table~\ref{tab:s1100}, but for Run 4 at 200 Myrs.}}} \\
\end{tabular}
\end{table*}

\begin{table*}[!ht]
\caption{Run 5 at 100 Myrs}
\hspace{-1.2cm}
\label{tab:my_complex_table_run5_100Myr}
\begin{tabular}{l l|*{5}{c}|c}
\hline
\multicolumn{2}{c|}{\multirow{2}{*}{Sight Lines}} & \multicolumn{5}{c|}{Results} & \multirow{2}{*}{$\log N(O~VI)_{sim}$} \\ 
\cline{3-7}
& & v$_{c}$ (km s$^{-1}$) & $b$ (km s$^{-1}$) & $b_T$ (km s$^{-1}$) & $b_{nt}$ (km s$^{-1}$) & $\log N(O~VI)$ \\ 
\hline
y & y1 
& -2.1$\pm$0.1 & 6.7$\pm$0.1 &  2.0 & 6.4 & 13.68$\pm$0.01 & 13.94 \\ 
& y2 
& -0.2$\pm$0.1 & 7.9$\pm$0.1 & 1.9  & 7.7 & 13.78$\pm$0.01 &  13.97 \\ 
& y3  
& 2.7$\pm$0.1 & 10.9$\pm$0.1 & 2.1  & 10.7 & 14.36$\pm$0.01 &  14.59 \\
\hline
z & 
z1  
& 2.5$\pm$0.1 & 6.9$\pm$0.1 & 4.4  & 5.3 & 13.68$\pm$0.01 &  13.86 \\ 
& z2  
& 5.7$\pm$0.1 & 7.5$\pm$0.1 &  3.8 & 6.5 & 13.71$\pm$0.01 & 13.95 \\
& z3  
& 3.1$\pm$0.1  & 7.1$\pm$0.1 & 3.3  & 6.3 & 13.93$\pm$0.01 &  14.15 \\
\hline
oblique & oblique1  
& -2.9$\pm$0.1 & 6.6$\pm$0.1 &  2.1 & 6.3 & 13.68$\pm$0.01 &  13.77 \\
& oblique2  
& -3.2$\pm$0.1 & 7.6$\pm$0.1 &  2.1  & 7.3 &  13.88$\pm$0.01 & 13.96 \\
& oblique3  
& -3.9$\pm$0.1 & 7.9$\pm$0.1 & 2.7  & 7.4 & 13.90$\pm$0.01 &  13.94 \\
\hline 
\multicolumn{8}{@{}l}{\parbox{0.9\textwidth}{\small\vspace{1mm}{Same as Table~\ref{tab:s1100}, but for Run 5 at 100 Myrs.}}} \\
\end{tabular}
\end{table*}

\begin{table*}[!ht]
\caption{Run 5 at 150 Myrs}
\hspace{-1.2cm}
\label{tab:my_complex_table_run5_150Myr}
\begin{tabular}{l l|*{5}{c}|c}
\hline
\multicolumn{2}{c|}{\multirow{2}{*}{Sight Lines}} & \multicolumn{5}{c|}{Results} & \multirow{2}{*}{$\log N(O~VI)_{sim}$} \\ 
\cline{3-7}
& & v$_{c}$ (km s$^{-1}$) & $b$ (km s$^{-1}$) & $b_T$ (km s$^{-1}$) & $b_{nt}$ (km s$^{-1}$) & $\log N(O~VI)$ \\ 
\hline
y & y1 
& -2.1$\pm$0.1 & 6.8$\pm$0.1 &  2.0 & 6.5 & 13.66$\pm$0.01 &  13.89 \\ 
& y2 
& -6.0$\pm$0.1 & 6.9$\pm$0.1 & 2.1  & 6.6 & 13.78$\pm$0.01 & 13.95  \\ 
& y3  
& -1.2$\pm$0.1 & 9.4$\pm$0.1 &  2.2 & 9.1 & 14.23$\pm$0.01 & 14.54 \\
\hline
z & 
z1  
& 10.8$\pm$0.1 & 6.3$\pm$0.1 & 3.3  & 5.4 & 13.78$\pm$0.01 & 14.17 \\ 
&   
& 17.4$\pm$0.1 & 18.4$\pm$0.1 &  9.7  & 15.6 & 13.46$\pm$0.01 &  \\
& z2  
& 16.6$\pm$0.1 & 6.4$\pm$0.1 &  3.6 & 5.3 & 13.77$\pm$0.01 &  14.12 \\ 
&  
& 18.3$\pm$0.1 & 15.5$\pm$0.1 &  12.0 & 9.8 & 13.67$\pm$0.01 & \\
& z3  
&  16.0$\pm$0.1 & 7.5$\pm$0.1 &  5.2 & 5.4 & 13.98$\pm$0.01 &  14.30 \\ 
& 
& 21.5$\pm$0.1 & 19.6$\pm$0.2 & 5.4 & 18.8 & 13.72$\pm$0.01 & \\ 
\hline
oblique & oblique1  
& -4.6$\pm$0.1 & 6.7$\pm$0.1 & 2.0  & 6.4 & 13.71$\pm$0.01 &  13.90 \\ 
& oblique2  
& 5.7$\pm$0.1 & 8.1$\pm$0.1 &  2.3  & 7.8 & 13.84$\pm$0.01  &  13.87 \\
& oblique3  
& 1.9$\pm$0.1 & 10.8$\pm$0.1 &  2.2 & 10.6 & 14.36$\pm$0.01 &  14.56 \\
\hline 
\multicolumn{8}{@{}l}{\parbox{0.9\textwidth}{\small\vspace{1mm}{Same as Table~\ref{tab:s1100}, but for Run 5 at 150 Myrs.}}} \\
\end{tabular}
\end{table*}

\begin{table*}[!ht]
\caption{Run 5 at 200 Myrs}
\hspace{-1.2cm}
\label{tab:my_complex_table_run5_200Myr}
\begin{tabular}{l l|*{5}{c}|c}
\hline
\multicolumn{2}{c|}{\multirow{2}{*}{Sight Lines}} & \multicolumn{5}{c|}{Results} & \multirow{2}{*}{$\log N(O~VI)_{sim}$} \\ 
\cline{3-7}
& & v$_{c}$ (km s$^{-1}$) & $b$ (km s$^{-1}$) & $b_T$ (km s$^{-1}$) & $b_{nt}$ (km s$^{-1}$) & $\log N(O~VI)$ \\ 
\hline
y & y1 
& -9.9$\pm$0.1 & 9.0$\pm$0.1 & 3.3  & 8.4 & 13.62$\pm$0.01 & 13.56 \\
& y2 
& -1.6$\pm$0.1 & 6.9$\pm$0.1 & 2.1  & 6.6 & 13.59$\pm$0.01 & 13.78 \\ 
& y3  
& 5.1$\pm$0.3 & 19.1$\pm$0.5 & 2.2  & 19.0 & 14.44$\pm$0.01 &  14.35 \\
\hline
z & 
z1  
& 16.1$\pm$0.2 & 15.8$\pm$0.3 & 3.9  & 15.3 & 14.63$\pm$0.01 &  14.86 \\ 
&   
& 34.0$\pm$0.4 & 10.2$\pm$0.5 & 4.1   & 9.3 & 13.96$\pm$0.03 &  \\ 
& z2  
& 33.5$\pm$0.2 & 15.2$\pm$0.4 & 4.6  & 14.5 & 14.28$\pm$0.01 & 14.30 \\ 
& z3  
& 37.9$\pm$0.1  & 10.6$\pm$0.1 & 5.7  & 8.9 & 14.04$\pm$0.01 & 14.06 \\
\hline
oblique & oblique1  
& 4.1$\pm$0.1 & 7.7$\pm$0.1 &  2.4 & 7.3 & 13.62$\pm$0.01 &  13.59 \\
& oblique2  
& 3.1$\pm$0.1 & 7.1$\pm$0.1 &  2.1  & 6.8 & 13.71$\pm$0.01  &  13.87 \\ 
& oblique3  
& 12.3$\pm$0.1 & 11.5$\pm$0.1 &  2.2 & 11.3 & 13.85$\pm$0.01 &  13.76 \\ 
\hline 
\multicolumn{8}{@{}l}{\parbox{0.9\textwidth}{\small\vspace{1mm}{Same as Table~\ref{tab:s1100}, but for Run 5 at 200 Myrs.}}} \\
\end{tabular}
\end{table*}

\begin{table*}[!ht]
\caption{Run 6 at 100 Myrs}
\hspace{-1.2cm}
\label{tab:my_complex_table_run6_100Myr}
\begin{tabular}{l l|*{5}{c}|c}
\hline
\multicolumn{2}{c|}{\multirow{2}{*}{Sight Lines}} & \multicolumn{5}{c|}{Results} & \multirow{2}{*}{$\log N(O~VI)_{sim}$} \\ 
\cline{3-7}
& & v$_{c}$ (km s$^{-1}$) & $b$ (km s$^{-1}$) & $b_T$ (km s$^{-1}$) & $b_{nt}$ (km s$^{-1}$) & $\log N(O~VI)$ \\ 
\hline
y & y1 
& -0.5$\pm$0.1 & 9.5$\pm$0.1 & 2.3  & 9.2 & 14.30$\pm$0.01 &  14.97 \\
& y2 
& -11.6$\pm$0.2 & 13.0$\pm$0.2 & 2.1  & 12.8 & 14.95$\pm$0.06 & 14.92 \\ 
& y3  
& -22.0$\pm$0.2 & 9.2$\pm$0.4 &  3.8 & 8.4 & 13.43$\pm$0.02 &  14.74 \\ 
&   
& -3.0$\pm$0.1 & 10.2$\pm$0.2 &  2.1 & 10.0 & 14.39$\pm$0.01 &  \\
\hline
z & 
z1  
& 21.0$\pm$0.9 & 67.0$\pm$1.1 & 16.8  & 64.9 & 14.23$\pm$0.01 & 15.43 \\ 
&   
& 26.9$\pm$0.1 & 14.3$\pm$0.2 & 4.4   & 13.6 & 14.78$\pm$0.01 &  \\ 
&   
& 56.7$\pm$0.1 & 16.7$\pm$0.2 &  5.4 & 15.8 & 14.83$\pm$0.01 &  \\ 
& z2  
& 36.2$\pm$0.1 & 16.6$\pm$0.2 &  2.8 & 16.4 & 14.93$\pm$0.02 & 15.34 \\ 
&  
& 64.8$\pm$0.1 & 9.2$\pm$0.2 & 6.0  & 7.0 & 14.20$\pm$0.01 & \\
& z3  
& 49.4$\pm$0.1  & 16.9$\pm$0.3 & 2.4  & 16.7 & 15.30$\pm$0.02 &  15.15 \\
\hline
oblique & oblique1  
& 12.8$\pm$0.1 & 13.9$\pm$0.2 & 1.9  & 13.8 & 15.02$\pm$0.02 & 15.07 \\
& oblique2  
& 19.7$\pm$0.1 & 14.3$\pm$0.3 &  2.2  & 14.1 & 14.98$\pm$0.03  & 15.03 \\
& oblique3  
& 27.5$\pm$3.2 & 15.1$\pm$1.9 & 2.4  & 14.9 & 14.83$\pm$0.06 & 14.98 \\ 
&  
& 46.1$\pm$3.0 & 10.9$\pm$1.6 &  2.4  & 10.6 & 14.50$\pm$0.27 & \\
\hline 
\multicolumn{8}{@{}l}{\parbox{0.9\textwidth}{\small\vspace{1mm}{Same as Table~\ref{tab:s1100}, but for Run 6 at 100 Myrs.}}} \\
\end{tabular}
\end{table*}

\begin{table*}[!ht]
\caption{Run 6 at 150 Myrs}
\hspace{-1.2cm}
\label{tab:my_complex_table_run6_150Myr}
\begin{tabular}{l l|*{5}{c}|c}
\hline
\multicolumn{2}{c|}{\multirow{2}{*}{Sight Lines}} & \multicolumn{5}{c|}{Results} & \multirow{2}{*}{$\log N(O~VI)_{sim}$} \\ 
\cline{3-7}
& & v$_{c}$ (km s$^{-1}$) & $b$ (km s$^{-1}$) & $b_T$ (km s$^{-1}$) & $b_{nt}$ (km s$^{-1}$) & $\log N(O~VI)$ \\ 
\hline
y & y1 
& -8.6$\pm$1.4 & 12.6$\pm$0.6 & 2.2 & 12.4 & 13.79$\pm$0.12 & 14.73 \\ 
& 
& -3.2$\pm$0.1 & 9.2$\pm$0.2 & 2.2 & 8.9 & 14.14$\pm$0.06 & \\
& y2 
& -10.5$\pm$0.1 & 11.0$\pm$0.1 & 2.1  & 10.8 & 14.42$\pm$0.01 & 14.82 \\ 
& y3  
& -0.4$\pm$0.1 & 8.3$\pm$0.1 & 2.2  & 8.0 & 14.04$\pm$0.01 & 14.61 \\ 
\hline
z & 
z1  
& 59.8$\pm$0.1 & 27.6$\pm$0.6 & 2.2  & 27.5 & 16.22$\pm$0.15 & 16.07 \\
& z2  
& 78.3$\pm$0.1 & 16.6$\pm$0.4 & 2.6  & 16.4 & 14.87$\pm$0.03 & 14.87 \\
& z3  
&  90.3$\pm$0.1 & 11.9$\pm$0.2 &  2.2 & 11.7 & 14.40$\pm$0.01 & 14.43 \\
\hline
oblique & oblique1  
& 22.3$\pm$0.1 & 13.1$\pm$0.1 & 2.1  & 12.9 & 14.59$\pm$0.01 &  14.89 \\ 
&  
& 47.7$\pm$0.3 & 11.4$\pm$0.6 & 5.1  & 10.2 & 13.24$\pm$0.02 & \\
& oblique2  
& 17.0$\pm$0.1 & 9.9$\pm$0.1 &  2.1  & 9.7 &  14.33$\pm$0.01 & 14.88 \\ 
& 
& 37.0$\pm$0.1 & 9.4$\pm$0.1 &  2.3 & 9.1 & 14.17$\pm$0.01 & \\
& oblique3  
& -3.2$\pm$0.3 & 7.3$\pm$0.6 & 2.2  & 7.0 & 13.95$\pm$0.07 & 14.42 \\ 
&  
& 4.2$\pm$1.6 & 16.7$\pm$1.3 &  2.3  & 16.5 & 13.88$\pm$0.08 & \\
\hline 
\multicolumn{8}{@{}l}{\parbox{0.9\textwidth}{\small\vspace{1mm}{Same as Table~\ref{tab:s1100}, but for Run 6 at 150 Myrs.}}} \\
\end{tabular}
\end{table*}

\begin{table*}[!ht]
\caption{Run 6 at 200 Myrs}
\hspace{-1.2cm}
\label{tab:my_complex_table_run6_200Myr}
\begin{tabular}{l l|*{5}{c}|c}
\hline
\multicolumn{2}{c|}{\multirow{2}{*}{Sight Lines}} & \multicolumn{5}{c|}{Results} & \multirow{2}{*}{$\log N(O~VI)_{sim}$} \\ 
\cline{3-7}
& & v$_{c}$ (km s$^{-1}$) & $b$ (km s$^{-1}$) & $b_T$ (km s$^{-1}$) & $b_{nt}$ (km s$^{-1}$) & $\log N(O~VI)$ \\ 
\hline
y & y1 
& -9.6$\pm$3.3 & 14.7$\pm$2.2 &  1.9 & 14.6 & 13.78$\pm$0.20 & 14.68 \\ 
&   
& -0.9$\pm$0.3 & 8.6$\pm$0.8 &  1.9 & 8.4 & 14.15$\pm$0.08 &  \\ 
& y2 
& -6.9$\pm$0.1 & 13.4$\pm$0.1 & 2.1  & 13.2 & 13.69$\pm$0.01 &  14.71 \\ 
& 
& -2.6$\pm$0.1 & 7.3$\pm$0.1 & 2.1 & 7.0 & 14.02$\pm$0.01 & \\
& y3  
& 1.8$\pm$0.1 & 11.0$\pm$0.1 & 1.9  & 10.8 & 14.40$\pm$0.01 & 14.62 \\
\hline
z & 
z1  
& 65.7$\pm$0.2 & 25.8$\pm$0.7 & 2.0  & 25.7 & 16.23$\pm$0.21 &  16.08 \\ 
& z2  
& 65.7$\pm$0.2 & 28.5$\pm$0.8 & 2.3  & 28.4 & 15.85$\pm$0.16 & 15.70 \\
& z3  
&  54.3$\pm$0.7 & 16.1$\pm$0.9 &  2.1 & 16.0 & 14.86$\pm$0.05 & 15.25 \\ 
& 
& 81.9$\pm$0.8 & 16.9$\pm$0.8 & 2.2 & 16.8 & 14.87$\pm$0.05 & \\ 
\hline
oblique & oblique1  
& 13.5$\pm$0.1 & 14.0$\pm$0.2 &  2.3 & 13.8 & 14.89$\pm$0.02 & 15.06 \\
& oblique2  
& 27.7$\pm$0.1 & 15.2$\pm$0.1 &  2.1  & 15.1 & 14.76$\pm$0.01  & 14.73 \\
& oblique3  
& 25.7$\pm$0.1 & 11.4$\pm$0.1 & 2.1  & 11.2 & 14.53$\pm$0.01 & 14.85 \\
\hline 
\multicolumn{8}{@{}l}{\parbox{0.9\textwidth}{\small\vspace{1mm}{Same as Table~\ref{tab:s1100}, but for Run 6 at 200 Myrs.}}} \\
\end{tabular}
\end{table*}

\begin{table*}[!ht]
\caption{Run 7 at 100 Myrs}
\hspace{-1.2cm}
\label{tab:my_complex_table_run7_100Myr}
\begin{tabular}{l l|*{5}{c}|c}
\hline
\multicolumn{2}{c|}{\multirow{2}{*}{Sight Lines}} & \multicolumn{5}{c|}{Results} & \multirow{2}{*}{$\log N(O~VI)_{sim}$} \\ 
\cline{3-7}
& & v$_{c}$ (km s$^{-1}$) & $b$ (km s$^{-1}$) & $b_T$ (km s$^{-1}$) & $b_{nt}$ (km s$^{-1}$) & $\log N(O~VI)$ \\ 
\hline
y & y1 
& -6.5$\pm$0.5 & 28.8$\pm$0.6 & 19.1  & 21.6 & 13.27$\pm$0.02 & 15.22 \\ 
&   
& 1.1$\pm$0.1 & 8.4$\pm$0.1 &  2.3 & 8.1 & 14.19$\pm$0.01 &  \\ 
& y2 
& -23.3$\pm$0.6 & 8.5$\pm$1.0 & 4.9  & 6.9 & 12.95$\pm$0.04 & 15.12 \\ 
& 
& 2.4$\pm$0.1 & 12.5$\pm$0.2 & 1.7 & 12.4 & 14.55$\pm$0.01 & \\
& y3  
& -8.7$\pm$0.1 & 17.0$\pm$0.4 & 1.7  & 16.9 & 15.53$\pm$0.02 & 15.38 \\
\hline
z & 
z1  
& -23.1$\pm$1.3 & 72.4$\pm$3.9 & 31.6  & 65.1 & 13.78$\pm$0.03 & 15.86 \\ 
&   
& -18.6$\pm$0.1 & 21.3$\pm$0.2 & 2.9   & 21.1 & 15.75$\pm$0.02 &  \\
& z2  
& -25.0$\pm$3.3 & 71.9$\pm$11.8 & 19.8  & 69.1 & 13.80$\pm$0.09 & 15.35 \\ 
&  
& 8.4$\pm$0.3 & 18.6$\pm$0.5 & 2.5  & 18.4 & 14.80$\pm$0.02 & \\
& z3  
&  -19.0$\pm$1.6 & 17.2$\pm$3.4 &  2.5 & 17.0 & 12.80$\pm$0.07 & 15.52 \\ 
& 
& 15.6$\pm$0.1 & 17.3$\pm$0.2 & 2.8 & 17.1 & 14.93$\pm$0.01 & \\ 
& 
& 47.0$\pm$0.4 & 13.0$\pm$0.8 & 6.6 & 11.2 & 13.22$\pm$0.03 & \\
\hline
oblique & oblique1  
& -1.5$\pm$0.1 & 8.8$\pm$0.1 & 2.0  & 8.6 & 14.15$\pm$0.01 & 14.96 \\ 
& 
& 12.4$\pm$1.6  & 40.8$\pm$2.4 & 22.0 & 34.4 & 12.95$\pm$0.03 & \\
& oblique2  
& -12.3$\pm$0.1 & 17.3$\pm$0.1 &  1.7  & 17.2 &  14.92$\pm$0.01 & 15.30 \\
& oblique3  
& -16.2$\pm$0.5 & 14.8$\pm$0.6 & 1.6  & 14.7 & 14.70$\pm$0.02 & 15.35 \\ 
&  
& 4.4$\pm$0.5 & 12.8$\pm$0.6 &  1.6  & 12.7 & 14.46$\pm$0.03 & \\
\hline 
\multicolumn{8}{@{}l}{\parbox{0.9\textwidth}{\small\vspace{1mm}{Same as Table~\ref{tab:s1100}, but for Run 7 at 100 Myrs.}}} \\
\end{tabular}
\end{table*}

\begin{table*}[!ht]
\caption{Run 7 at 150 Myrs}
\hspace{-1.2cm}
\label{tab:my_complex_table_run7_150Myr}
\begin{tabular}{l l|*{5}{c}|c}
\hline
\multicolumn{2}{c|}{\multirow{2}{*}{Sight Lines}} & \multicolumn{5}{c|}{Results} & \multirow{2}{*}{$\log N(O~VI)_{sim}$} \\ 
\cline{3-7}
& & v$_{c}$ (km s$^{-1}$) & $b$ (km s$^{-1}$) & $b_T$ (km s$^{-1}$) & $b_{nt}$ (km s$^{-1}$) & $\log N(O~VI)$ \\ 
\hline
y & y1 
& -5.1$\pm$2.8 & 15.1$\pm$1.0 & 4.7  & 14.3 & 13.23$\pm$0.22 & 15.41 \\ 
&   
& 0.4$\pm$0.1 & 8.3$\pm$0.3 &  1.8 & 8.1 & 14.19$\pm$0.02 &  \\ 
& y2 
& -25.1$\pm$0.5 & 6.5$\pm$0.6 & 3.4  & 5.5 & 13.00$\pm$0.02 & 15.17 \\ 
& 
& 2.0$\pm$0.1 & 13.8$\pm$0.1 & 1.5 & 13.7 & 14.76$\pm$0.01 & \\
& y3  
& -120.5$\pm$5.0 & 38.4$\pm$10.9 & 16.3  & 34.8 & 13.26$\pm$0.10 & 14.79 \\ 
&   
& -73.1$\pm$0.6 & 16.6$\pm$1.4  & 13.0  & 10.3 & 13.67$\pm$0.03 &  \\ 
&   
& -47.4$\pm$0.5 & 7.0$\pm$0.6 & 2.7  & 6.5 & 13.33$\pm$0.04 &  \\ 
& 
& -3.9$\pm$0.1 & 9.6$\pm$0.3 & 2.0 & 9.4 & 14.12$\pm$0.01 & \\
\hline
z & 
z1  
& -27.7$\pm$8.1 & 77.0$\pm$13.5 & 30.7  & 70.6 & 13.87$\pm$0.11 & 15.82 \\ 
&   
& -14.6$\pm$0.2 & 23.2$\pm$0.5 &  2.5  & 23.1 & 15.16$\pm$0.02 &  \\ 
&   
& 78.8$\pm$2.4 & 44.4$\pm$4.1 &  3.7 & 44.2 & 13.98$\pm$0.04 &  \\ 
& 
& 94.7$\pm$0.2 & 9.0$\pm$0.2 & 3.0 & 8.5 & 14.05$\pm$0.02 & \\
& z2  
& -103.1$\pm$4.4 & 73.3$\pm$5.9 &  20.0 & 70.5 & 13.97$\pm$0.05 & 15.70 \\ 
&  
& -9.6$\pm$0.3 & 24.0$\pm$0.7 &  2.0 & 23.9 & 15.21$\pm$0.05 & \\ 
& 
& 42.4$\pm$1.5 & 9.1$\pm$2.8 & 4.1 & 8.1 & 13.09$\pm$0.11 & \\
& z3  
& -26.7$\pm$1.5  & 14.9$\pm$2.9 &  13.0 & 7.3 & 13.45$\pm$0.07 & 15.44 \\ 
& 
& 20.3$\pm$0.3 & 22.5$\pm$0.5 & 2.4 & 22.4 & 14.95$\pm$0.02 & \\
\hline
oblique & oblique1  
& -4.7$\pm$0.1 & 11.5$\pm$0.2 & 1.5  & 11.4 & 14.66$\pm$0.02 & 15.63 \\
& oblique2  
& -16.5$\pm$0.1 & 11.2$\pm$0.1 &  1.8  & 11.1 & 14.58$\pm$0.01  & 15.30 \\ 
& 
& 3.5$\pm$0.1 & 8.8$\pm$0.1 &  1.7 & 8.6 & 14.23$\pm$0.01 & \\
& oblique3  
& 5.8$\pm$0.3 & 15.4$\pm$1.0 &  2.2 & 15.2 & 15.17$\pm$0.08 & 15.22 \\ 
&  
& 46.0$\pm$5.8 & 53.0$\pm$9.7 &  21.1  & 48.6 & 13.99$\pm$0.08  & \\
\hline 
\multicolumn{8}{@{}l}{\parbox{0.9\textwidth}{\small\vspace{1mm}{Same as Table~\ref{tab:s1100}, but for Run 7 at 150 Myrs.}}} \\
\end{tabular}
\end{table*}

\begin{table*}[!ht]
\caption{Run 7 at 200 Myrs}
\hspace{-1.2cm}
\label{tab:my_complex_table_run7_200Myr}
\begin{tabular}{l l|*{5}{c}|c}
\hline
\multicolumn{2}{c|}{\multirow{2}{*}{Sight Lines}} & \multicolumn{5}{c|}{Results} & \multirow{2}{*}{$\log N(O~VI)_{sim}$} \\ 
\cline{3-7}
& & v$_{c}$ (km s$^{-1}$) & $b$ (km s$^{-1}$) & $b_T$ (km s$^{-1}$) & $b_{nt}$ (km s$^{-1}$) & $\log N(O~VI)$ \\ 
\hline
y & y1 
& 1.0$\pm$0.1 & 10.4$\pm$0.1 & 2.1  & 10.2 & 14.36$\pm$0.01 & 15.34 \\
& y2 
& -2.5$\pm$0.2 & 11.3$\pm$0.4 & 1.6  & 11.2 & 14.34$\pm$0.02 & 15.34 \\ 
& y3  
& -2.2$\pm$0.1 & 11.5$\pm$0.1 &  1.7 & 11.4 & 14.56$\pm$0.01 & 14.94 \\ 
&   
& 13.6$\pm$0.3 & 15.2$\pm$0.3 & 1.7  & 15.1 & 14.09$\pm$0.01 &  \\
\hline
z & 
z1  
& -67.2$\pm$8.1 & 44.9$\pm$0.1 &  23.6 & 38.2 & 13.65$\pm$0.14 & 15.85 \\ 
&   
& -12.9$\pm$0.4 & 17.5$\pm$1.2 &  2.2  & 17.4 & 15.13$\pm$0.10 &  \\ 
&   
& 54.6$\pm$1.1 & 43.8$\pm$2.2 &  3.1 & 43.7 & 14.65$\pm$0.02 &  \\ 
& z2  
& 24.5$\pm$0.1 & 19.4$\pm$0.3 & 3.0  & 19.2 & 15.48$\pm$0.03 & 15.73 \\ 
&  
& 27.6$\pm$1.1 & 55.0$\pm$4.2 & 16.1  & 52.6 & 13.99$\pm$0.05 & \\
& z3  
& 31.9$\pm$0.7  & 12.1$\pm$1.4 & 2.5  & 11.8 & 14.49$\pm$0.08 & 15.07 \\ 
& 
& 42.7$\pm$2.0 & 23.7$\pm$1.4  & 3.0 & 23.5 & 14.46$\pm$0.08 & \\
\hline
oblique & oblique1  
& -5.5$\pm$0.1 & 9.5$\pm$0.1 &  1.8 & 9.3 & 14.31$\pm$0.01 & 15.24 \\ 
&  
& 7.5$\pm$0.1 & 8.1$\pm$0.1 & 1.8  & 7.9 & 13.57$\pm$0.01 & \\
& oblique2  
& -3.5$\pm$0.1 & 15.5$\pm$0.2 &  1.6  & 15.4 & 15.09$\pm$0.02  & 15.32 \\
& oblique3  
& 35.8$\pm$0.2 & 8.0$\pm$0.6 & 2.4  & 7.6 & 13.93$\pm$0.02 & 14.94 \\ 
&  
& 83.8$\pm$0.2 & 11.6$\pm$0.5 &  2.5  & 11.3 & 14.39$\pm$0.02 &  \\ 
& 
& 118.7$\pm$35.3 & 60.9$\pm$18.8 & 23.6 & 56.1 & 13.49$\pm$0.13 & \\
\hline 
\multicolumn{8}{@{}l}{\parbox{0.9\textwidth}{\small\vspace{1mm}{Same as Table~\ref{tab:s1100}, but for Run 7 at 200 Myrs.}}} \\
\end{tabular}
\end{table*}

\begin{table*}[!ht]
\caption{Run 8 at 100 Myrs}
\hspace{-1.2cm}
\label{tab:my_complex_table_run8_100Myr}
\begin{tabular}{l l|*{5}{c}|c}
\hline
\multicolumn{2}{c|}{\multirow{2}{*}{Sight Lines}} & \multicolumn{5}{c|}{Results} & \multirow{2}{*}{$\log N(O~VI)_{sim}$} \\ 
\cline{3-7}
& & v$_{c}$ (km s$^{-1}$) & $b$ (km s$^{-1}$) & $b_T$ (km s$^{-1}$) & $b_{nt}$ (km s$^{-1}$) & $\log N(O~VI)$ \\ 
\hline
y & y1 
& 3.6$\pm$0.1 & 10.1$\pm$0.1 & 2.2  & 9.9 & 14.36$\pm$0.01 & 15.10 \\ 
&   
& 14.2$\pm$2.6 & 52.6$\pm$4.9 &  28.5 & 44.2 & 12.86$\pm$0.04 &  \\ 
& y2 
& -11.9$\pm$0.1 & 13.2$\pm$0.2 &  2.2 & 13.0 & 14.69$\pm$0.02 & 14.70 \\ 
& y3  
& -9.3$\pm$0.2 & 13.8$\pm$0.3 & 2.0  & 13.7 & 14.50$\pm$0.02 & 14.95 \\  
\hline
z & 
z1  
& 10.6$\pm$0.5 & 26.9$\pm$1.0 &  2.6 & 26.8 & 14.86$\pm$0.03 & 15.79 \\ 
&   
& 67.4$\pm$0.4 & 20.1$\pm$0.9 &  2.7  & 19.9 & 15.03$\pm$0.06 &  \\
& z2  
& 7.6$\pm$0.3 & 17.0$\pm$0.8 & 1.8  & 16.9 & 14.89$\pm$0.06 & 15.38 \\ 
&  
& 52.5$\pm$0.5 & 21.6$\pm$0.9 & 3.7  & 21.3 & 14.55$\pm$0.02 & \\
& z3  
& 20.7$\pm$0.3  & 18.2$\pm$1.0 & 2.5  & 18.0 & 15.22$\pm$0.09 & 15.40 \\ 
& 
& 69.8$\pm$1.5 & 25.6$\pm$3.2 & 8.5 & 24.1 & 13.95$\pm$0.05 & \\
\hline
oblique & oblique1  
& 8.6$\pm$0.6 & 18.6$\pm$1.5 & 2.1  & 18.5 & 14.61$\pm$0.04 & 15.58 \\ 
&  
& 37.3$\pm$2.3 & 14.3$\pm$5.5 &  2.2 & 14.1 & 13.63$\pm$0.15 & \\ 
& 
& 57.3$\pm$1.3 & 9.4$\pm$3.3 & 2.6 & 9.0 & 13.43$\pm$0.12 & \\
& oblique2  
& 8.5$\pm$0.1 & 13.2$\pm$0.2 &  2.2  & 13.0 & 14.37$\pm$0.01  & 14.64 \\ 
& 
& 26.5$\pm$0.1 & 9.2$\pm$0.2 & 2.2  & 8.9 & 14.20$\pm$0.01 & \\
& oblique3  
& 23.6$\pm$0.2 & 10.5$\pm$0.4 & 1.8  & 10.3 & 14.27$\pm$0.02 & 14.96 \\ 
\hline 
\multicolumn{8}{@{}l}{\parbox{0.9\textwidth}{\small\vspace{1mm}{Same as Table~\ref{tab:s1100}, but for Run 8 at 100 Myrs.}}} \\
\end{tabular}
\end{table*}

\begin{table*}[!ht]
\caption{Run 8 at 150 Myrs}
\hspace{-1.2cm}
\label{tab:my_complex_table_run8_150Myr}
\begin{tabular}{l l|*{5}{c}|c}
\hline
\multicolumn{2}{c|}{\multirow{2}{*}{Sight Lines}} & \multicolumn{5}{c|}{Results} & \multirow{2}{*}{$\log N(O~VI)_{sim}$} \\ 
\cline{3-7}
& & v$_{c}$ (km s$^{-1}$) & $b$ (km s$^{-1}$) & $b_T$ (km s$^{-1}$) & $b_{nt}$ (km s$^{-1}$) & $\log N(O~VI)$ \\ 
\hline
y & y1 
& -3.4$\pm$0.2 & 12.8$\pm$0.3 & 1.6  & 12.7 & 14.49$\pm$0.02 & 14.72 \\ 
& y2 
& -30.1$\pm$2.8 & 45.5$\pm$4.8 &  3.2 & 45.4 & 13.26$\pm$0.04 & 15.39 \\ 
& 
& -2.5$\pm$0.1 & 9.3$\pm$0.1 & 2.0 & 9.1 & 14.16$\pm$0.01 & \\
& y3  
& 1.5$\pm$0.1 & 8.5$\pm$0.1 & 1.9  & 8.3 & 14.17$\pm$0.01 & 14.93 \\
\hline
z & 
z1  
& 9.9$\pm$0.2 & 21.0$\pm$0.6 & 2.6  & 20.8 & 15.21$\pm$0.05 & 16.24 \\ 
&   
& 72.1$\pm$0.2 & 28.6$\pm$0.7 &  2.5  & 28.5 & 15.50$\pm$0.04 &  \\
& z2  
& 9.9$\pm$0.2 & 9.0$\pm$0.5 & 1.3  & 8.9 & 15.25$\pm$0.38 & 15.48 \\ 
&  
& 29.9$\pm$0.3 & 11.1$\pm$0.5 & 2.1  & 10.9 & 14.10$\pm$0.02 & \\ 
& 
& 60.4$\pm$0.2 & 11.6$\pm$0.3 & 4.2 & 10.8 & 14.05$\pm$0.01 & \\
& z3  
&  20.8$\pm$0.2 & 14.3$\pm$0.5 & 1.6  & 14.2 & 14.68$\pm$0.03 & 15.27  \\ 
& 
& 64.0$\pm$0.5 & 8.1$\pm$2.4 & 5.0 & 6.4 & 13.62$\pm$0.08 & \\ 
& 
& 82.4$\pm$3.0 & 43.3$\pm$4.2 & 4.7 & 43.0 & 13.97$\pm$0.05 & \\
\hline
oblique & oblique1  
& 8.2$\pm$0.9 & 14.7$\pm$1.9 & 2.2  & 14.5 & 14.36$\pm$0.15 & 14.64 \\ 
&  
& 22.6$\pm$4.4 & 25.0$\pm$3.4 &  2.3 & 24.9 & 14.36$\pm$0.15 & \\
& oblique2  
& 7.4$\pm$0.6 & 11.3$\pm$0.8 &  2.2  & 11.1 & 14.41$\pm$0.05  & 14.66 \\ 
& 
& 26.0$\pm$0.1 & 15.4$\pm$3.8 & 3.9  & 14.9 & 13.85$\pm$0.12 & \\
& oblique3  
& 29.3$\pm$0.1 & 11.7$\pm$0.2 & 1.9  & 11.5 & 14.26$\pm$0.01 & 14.40 \\ 
\hline 
\multicolumn{8}{@{}l}{\parbox{0.9\textwidth}{\small\vspace{1mm}{Same as Table~\ref{tab:s1100}, but for Run 8 at 150 Myrs.}}} \\
\end{tabular}
\end{table*}

\begin{table*}[!ht]
\caption{Run 8 at 200 Myrs}
\hspace{-1.2cm}
\label{tab:my_complex_table_run8_200Myr}
\begin{tabular}{l l|*{5}{c}|c}
\hline
\multicolumn{2}{c|}{\multirow{2}{*}{Sight Lines}} & \multicolumn{5}{c|}{Results} & \multirow{2}{*}{$\log N(O~VI)_{sim}$} \\ 
\cline{3-7}
& & v$_{c}$ (km s$^{-1}$) & $b$ (km s$^{-1}$) & $b_T$ (km s$^{-1}$) & $b_{nt}$ (km s$^{-1}$) & $\log N(O~VI)$ \\ 
\hline
y & y1 
& -11.4$\pm$5.3 & 20.6$\pm$4.4 &  10.0 & 18.0 & 13.01$\pm$0.20 & 15.11 \\ 
&   
& -0.4$\pm$0.1 & 9.2$\pm$0.2 &  2.7 & 8.8 & 14.24$\pm$0.01 &  \\ 
& y2 
& -7.2$\pm$0.1 & 17.3$\pm$0.3 & 2.0  & 17.2 & 14.86$\pm$0.02 & 14.87 \\ 
& y3  
& -54.7$\pm$1.8 & 32.2$\pm$3.9 & 9.1  & 30.9 & 13.70$\pm$0.04 & 14.60 \\ 
&   
& -8.3$\pm$0.2 & 15.7$\pm$0.4 & 2.2  & 15.5 & 14.55$\pm$0.02 &  \\
\hline
z & 
z1  
& 45.8$\pm$0.2 & 30.1$\pm$1.1 & 2.2  & 30.0 & 16.26$\pm$0.03 & 16.11 \\
& z2  
& 26.0$\pm$0.3 & 20.7$\pm$0.8 &  1.8 & 20.6 & 15.51$\pm$0.12 & 15.66 \\ 
&  
& 75.8$\pm$0.3 & 18.1$\pm$0.7 & 4.9  & 17.4 & 14.48$\pm$0.02 & \\
& z3  
& 30.1$\pm$0.4   & 16.6$\pm$0.8 & 1.8  & 16.5 & 14.79$\pm$0.04 & 15.29 \\ 
& 
& 65.4$\pm$5.1 & 35.1$\pm$0.3 & 2.0 & 35.0 & 13.79$\pm$0.11 & \\
\hline
oblique & oblique1  
& 4.9$\pm$0.1 & 14.2$\pm$0.1 & 1.7  & 14.1 & 14.73$\pm$0.01 & 15.30 \\
& oblique2  
& 15.5$\pm$0.2 & 18.8$\pm$0.4 &  2.2  & 18.7 &  14.85$\pm$0.02 & 15.10 \\
& oblique3  
& -7.1$\pm$0.1 & 10.2$\pm$0.1 &  2.0 & 10.0 & 14.25$\pm$0.01 & 14.94 \\ 
&  
& 12.6$\pm$3.7 & 35.2$\pm$5.0 &  15.6  & 31.6 & 13.25$\pm$0.08 & \\
\hline 
\multicolumn{8}{@{}l}{\parbox{0.9\textwidth}{\small\vspace{1mm}{Same as Table~\ref{tab:s1100}, but for Run 8 at 200 Myrs.}}} \\
\end{tabular}
\end{table*} 

\begin{table*}[!ht]
\caption{Run 9 at 100 Myrs}
\hspace{-1.2cm}
\label{tab:my_complex_table_run9_100Myr}
\begin{tabular}{l l|*{5}{c}|c}
\hline
\multicolumn{2}{c|}{\multirow{2}{*}{Sight Lines}} & \multicolumn{5}{c|}{Results} & \multirow{2}{*}{$\log N(O~VI)_{sim}$} \\ 
\cline{3-7}
& & v$_{c}$ (km s$^{-1}$) & $b$ (km s$^{-1}$) & $b_T$ (km s$^{-1}$) & $b_{nt}$ (km s$^{-1}$) & $\log N(O~VI)$ \\ 
\hline
y & y1 
& -7.7$\pm$0.1 & 13.5$\pm$0.2 &  2.1 & 13.3 & 14.86$\pm$0.02 & 15.39 \\
& y2 
& -17.2$\pm$3.3 & 16.7$\pm$2.6 & 2.2  & 16.6 & 14.54$\pm$0.17 & 15.01 \\ 
& 
& -4.3$\pm$1.7 & 11.2$\pm$1.3 & 2.2 & 11.0 & 14.49$\pm$0.21 & \\
& y3  
& -30.1$\pm$1.6 & 17.5$\pm$3.4 & 7.4  & 15.9 & 12.79$\pm$0.07 & 14.93 \\ 
&   
& -0.4$\pm$0.1 & 9.2$\pm$0.1 &  1.8 & 9.0 & 14.19$\pm$0.01 &  \\
\hline
z & 
z1  
& 33.2$\pm$1.3 & 20.0$\pm$1.1 &  2.3 & 19.9 & 15.57$\pm$0.09 & 16.20 \\ 
&   
& 84.5$\pm$2.4 & 30.6$\pm$2.5 & 3.5   & 30.4 & 15.35$\pm$0.06 &  \\ 
&   
& 120.1$\pm$6.5 & 91.4$\pm$5.8 & 28.6  & 86.8 & 14.67$\pm$0.06 &  \\ 
& z2  
& 35.0$\pm$0.5 & 9.0$\pm$2.2 & 1.9  & 8.8 & 15.27$\pm$0.97 & 15.67 \\ 
&  
& 52.7$\pm$1.5 & 29.9$\pm$1.4 &  2.3 & 29.8 & 14.54$\pm$0.04 & \\ 
& 
& 96.5$\pm$0.3 & 20.8$\pm$0.5 & 5.4 & 20.1 & 14.69$\pm$0.01 & \\ 
& 
& 173.5$\pm$12.9 & 96.7$\pm$27.7 & 34.4 & 90.4 & 13.52$\pm$0.10 & \\
& z3  
& 50.1$\pm$0.9 & 31.3$\pm$1.3 & 2.2  & 31.2 & 15.64$\pm$0.06 & 15.79 \\ 
& 
& 113.6$\pm$1.0 & 28.5$\pm$2.4 & 3.2 & 28.3 & 15.41$\pm$0.09 & \\ 
& 
& 163.5$\pm$0.7 & 14.0$\pm$1.3 & 6.8 & 12.2 & 14.00$\pm$0.04 & \\
\hline
oblique & oblique1  
& 4.7$\pm$0.3 & 9.7$\pm$0.7 &  2.0 & 9.5 & 14.23$\pm$0.03 & 15.21 \\ 
&  
& 34.3$\pm$0.5 & 17.8$\pm$0.8 &  1.9 & 17.7 & 14.48$\pm$0.02 & \\
& oblique2  
& 16.2$\pm$0.5 & 12.0$\pm$0.7 &  2.2  & 11.8 & 14.54$\pm$0.04  & 15.22 \\ 
& 
& 49.9$\pm$0.6 & 25.0$\pm$1.4 &  2.4 & 24.9 & 14.87$\pm$0.03 & \\
& oblique3  
& 31.5$\pm$0.1 & 8.2$\pm$0.1 & 1.8  & 8.0 & 14.17$\pm$0.01 & 15.00 \\ 
&  
& 59.3$\pm$0.1 & 9.0$\pm$0.3 &  3.5  & 8.3 & 13.18$\pm$0.01 & \\ 
& 
& 125.1$\pm$0.5 & 35.4$\pm$1.0 & 18.9 & 29.9 & 13.26$\pm$0.01 & \\
\hline 
\multicolumn{8}{@{}l}{\parbox{0.9\textwidth}{\small\vspace{1mm}{Same as Table~\ref{tab:s1100}, but for Run 9 at 100 Myrs.}}} \\
\end{tabular}
\end{table*}

\begin{table*}[!ht]
\caption{Run 9 at 150 Myrs}
\hspace{-1.2cm}
\label{tab:my_complex_table_run9_150Myr}
\begin{tabular}{l l|*{5}{c}|c}
\hline
\multicolumn{2}{c|}{\multirow{2}{*}{Sight Lines}} & \multicolumn{5}{c|}{Results} & \multirow{2}{*}{$\log N(O~VI)_{sim}$} \\ 
\cline{3-7}
& & v$_{c}$ (km s$^{-1}$) & $b$ (km s$^{-1}$) & $b_T$ (km s$^{-1}$) & $b_{nt}$ (km s$^{-1}$) & $\log N(O~VI)$ \\ 
\hline
y & y1 
& -0.6$\pm$0.1 & 15.4$\pm$0.3 &  2.5 & 15.2 & 15.28$\pm$0.03 & 15.44 \\
& y2 
& -30.8$\pm$0.7 & 11.7$\pm$1.0 & 3.0  & 11.3 & 13.50$\pm$0.05 & 14.94 \\ 
& 
& -13.0$\pm$0.1 & 11.7$\pm$0.2 & 2.2 & 11.5 & 14.53$\pm$0.01 & \\
& y3  
& -12.7$\pm$0.5 & 23.2$\pm$0.5 &  6.7 & 22.2 & 13.01$\pm$0.02 & 15.24 \\ 
&   
& -2.4$\pm$0.1 & 8.6$\pm$0.1 &  2.0 & 8.4 & 14.18$\pm$0.01 &  \\
\hline
z & 
z1  
& 56.0$\pm$0.3 & 30.5$\pm$1.0 & 2.7  & 30.4 & 15.76$\pm$0.08 & 16.29 \\ 
&   
& 143.0$\pm$0.4 & 24.6$\pm$0.6 &  3.0  & 24.4 & 15.12$\pm$0.02 &  \\ 
&   
& 185.4$\pm$2.5 & 21.4$\pm$4.5 &  15.6 & 14.6 & 13.69$\pm$0.09 &  \\ 
& z2  
& 57.0$\pm$0.3 & 17.9$\pm$0.7 &  2.9 & 17.7 & 15.15$\pm$0.06 & 15.30 \\ 
&  
& 115.4$\pm$0.5 & 38.9$\pm$1.0 & 6.2  & 38.4 & 14.89$\pm$0.01 & \\
& z3  
& 99.3$\pm$0.1  & 14.1$\pm$0.1 &  2.9 & 13.8 & 14.70$\pm$0.01 & 15.14 \\ 
& 
& 122.3$\pm$0.2 & 18.0$\pm$0.3 & 7.0 & 16.6 & 14.35$\pm$0.01 & \\ 
& 
& 162.2$\pm$2.2 & 46.5$\pm$4.0 & 30.7 & 34.9 & 13.37$\pm$0.04 & \\
\hline
oblique & oblique1  
& 13.2$\pm$0.1 & 13.7$\pm$0.2 & 2.1  & 13.5 & 14.79$\pm$0.01 & 15.45 \\ 
&  
& 57.6$\pm$4.1 & 24.6$\pm$8.5 & 11.6  & 22.0 & 12.71$\pm$0.13 & \\
& oblique2  
& 23.1$\pm$0.1 & 16.4$\pm$0.3 &  2.1  & 16.3 & 15.17$\pm$0.10  & 15.12 \\ 
& 
& 62.4$\pm$0.6 & 7.0$\pm$1.8 & 3.0  & 6.3 & 13.19$\pm$0.03 & \\
& oblique3  
& 19.8$\pm$0.2 & 9.9$\pm$0.4 &  2.3 & 9.6 & 14.36$\pm$0.03 & 15.36 \\ 
&  
& 52.5$\pm$0.4 & 29.4$\pm$0.8 &  3.2  & 29.2 & 14.92$\pm$0.01 & \\ 
& 
& 117.5$\pm$3.4 & 25.6$\pm$7.1 & 21.9 & 13.3 & 13.13$\pm$0.10 & \\
\hline 
\multicolumn{8}{@{}l}{\parbox{0.9\textwidth}{\small\vspace{1mm}{Same as Table~\ref{tab:s1100}, but for Run 9 at 150 Myrs.}}} \\
\end{tabular}
\end{table*}

\begin{table*}[!ht]
\caption{Run 9 at 200 Myrs}
\hspace{-1.2cm}
\label{tab:my_complex_table_run9_200Myr}
\begin{tabular}{l l|*{5}{c}|c}
\hline
\multicolumn{2}{c|}{\multirow{2}{*}{Sight Lines}} & \multicolumn{5}{c|}{Results} & \multirow{2}{*}{$\log N(O~VI)_{sim}$} \\ 
\cline{3-7}
& & v$_{c}$ (km s$^{-1}$) & $b$ (km s$^{-1}$) & $b_T$ (km s$^{-1}$) & $b_{nt}$ (km s$^{-1}$) & $\log N(O~VI)$ \\ 
\hline
y & y1 
& -33.0$\pm$1.6 & 8.2$\pm$7.1 & 4.2  & 7.0 & 12.70$\pm$0.17 & 15.09 \\ 
&   
& -4.2$\pm$0.1 & 12.3$\pm$0.4 &  2.0 & 12.1 & 14.59$\pm$0.03 &  \\ 
& y2 
& -31.5$\pm$0.8 & 9.0$\pm$1.1 & 4.2  & 8.0 & 13.34$\pm$0.07 & 14.92 \\ 
& 
& -11.4$\pm$0.2 & 12.6$\pm$0.4 & 2.1 & 12.4 & 14.59$\pm$0.02 & \\
& y3  
& -19.0$\pm$0.7 & 12.7$\pm$1.1 &  2.0 & 12.5 & 14.41$\pm$0.04 & 15.06 \\ 
&   
& -2.4$\pm$0.6 & 8.6$\pm$0.9 & 2.0  & 8.4 & 14.24$\pm$0.05 &  \\
\hline
z & 
z1  
& 35.9$\pm$0.7 & 13.4$\pm$0.5 & 2.1  & 13.2 & 14.67$\pm$0.04 & 16.63 \\ 
&   
& 120.0$\pm$5.5 & 45.7$\pm$2.0 &  2.9  & 45.6 & 15.63$\pm$0.07 &  \\ 
&   
& 201.1$\pm$0.9 & 19.6$\pm$1.7 & 7.3  & 18.2 & 13.25$\pm$0.04 &  \\ 
& z2  
& 59.7$\pm$1.0 & 22.5$\pm$0.5 &  2.4 & 22.4 & 15.92$\pm$0.07 & 16.13 \\ 
&  
& 111.3$\pm$0.6 & 20.6$\pm$0.9 &  4.7 & 20.1 & 15.98$\pm$0.11 & \\ 
& 
& 155.7$\pm$1.1 & 60.7$\pm$1.2 & 29.0 & 53.3 & 14.77$\pm$0.01 & \\
& z3  
& 52.7$\pm$0.1 & 18.1$\pm$0.4 &  2.8 & 17.9 & 15.42$\pm$0.05 & 15.67 \\ 
&  
& 94.8$\pm$0.5 & 13.1$\pm$3.2 & 5.4  & 11.9 & 14.82$\pm$0.06 & \\ 
& 
& 121.6$\pm$0.5 & 15.8$\pm$0.7 & 9.5 & 12.6 & 14.63$\pm$0.02 & \\ 
& 
& 162.3$\pm$1.7 & 51.2$\pm$3.1 & 23.5 & 45.5 & 14.18$\pm$0.02 & \\
\hline
oblique & oblique1  
& 13.5$\pm$0.1 & 10.7$\pm$0.1 &  1.9  & 10.5 & 14.41$\pm$0.01  & 15.15 \\ 
& 
& 21.1$\pm$0.1 & 7.1$\pm$0.1 &  1.9 & 6.8 & 13.94$\pm$0.01 & \\ 
&  
& 41.0$\pm$0.1 & 10.3$\pm$0.1 & 3.1  & 9.8 & 13.18$\pm$0.01 & \\
& oblique2  
& 27.4$\pm$0.1 & 12.6$\pm$0.1 &  1.9  & 12.5 &  14.56$\pm$0.01 & 14.98 \\ 
& 
& 54.6$\pm$0.1 & 13.9$\pm$0.2 &  3.1 & 13.5 & 14.39$\pm$0.01 & \\ 
&  
& 105.6$\pm$2.9 & 38.6$\pm$6.2 &  3.7 & 38.4 & 13.12$\pm$0.06 & \\
& oblique3  
& 46.1$\pm$0.1  & 14.2$\pm$0.1 & 2.1  & 14.0 & 14.72$\pm$0.01 & 15.45 \\
\hline 
\multicolumn{8}{@{}l}{\parbox{0.9\textwidth}{\small\vspace{1mm}{Same as Table~\ref{tab:s1100}, but for Run 9 at 200 Myrs.}}} \\
\end{tabular}
\end{table*} 

\begin{table*}[!ht]
\caption{Run 10 at 100 Myrs}
\hspace{-1.2cm}
\label{tab:my_complex_table_run10_100Myr}
\begin{tabular}{l l|*{5}{c}|c}
\hline
\multicolumn{2}{c|}{\multirow{2}{*}{Sight Lines}} & \multicolumn{5}{c|}{Results} & \multirow{2}{*}{$\log N(O~VI)_{sim}$} \\ 
\cline{3-7}
& & v$_{c}$ (km s$^{-1}$) & $b$ (km s$^{-1}$) & $b_T$ (km s$^{-1}$) & $b_{nt}$ (km s$^{-1}$) & $\log N(O~VI)$ \\ 
\hline
y & y1 
& -7.2$\pm$0.2 & 11.0$\pm$0.4 & 2.2  & 10.8 & 14.27$\pm$0.02 & 14.82 \\
& y2 
& -4.4$\pm$0.1 & 11.5$\pm$0.1 & 1.9  & 11.3 & 14.52$\pm$0.01 & 15.00 \\ 
& y3  
& -62.4$\pm$1.2 & 14.3$\pm$2.3 & 9.1  & 11.0 & 13.25$\pm$0.06 & 15.07 \\ 
&   
& -23.0$\pm$0.6 & 18.5$\pm$1.1 &  3.7 & 18.1 & 14.08$\pm$0.02 &  \\ 
&   
& -0.8$\pm$0.2 & 10.3$\pm$0.3 &  3.0 & 9.9 & 14.37$\pm$0.02 &  \\ 
\hline
z & 
z1  
& -20.3$\pm$2.6 & 102.3$\pm$4.2 & 25.3  & 99.1 & 14.43$\pm$0.03 & 15.79 \\ 
&   
& 4.0$\pm$0.2 & 28.9$\pm$0.8 &  4.0  & 28.6 & 15.64$\pm$0.01 &  \\
& z2  
& -73.6$\pm$1.7 & 92.1$\pm$3.6 &  21.4  & 89.6 & 13.89$\pm$0.01  & 15.29 \\ 
& 
& 21.0$\pm$0.1 & 11.6$\pm$0.1 &  2.1 & 11.4 & 14.53$\pm$0.01 & \\ 
&  
& 47.5$\pm$0.3 & 7.1$\pm$1.1 &  3.1 & 6.4 & 13.15$\pm$0.02 & \\
& z3  
& 35.7$\pm$0.2  & 18.4$\pm$0.4 & 3.3  & 18.1 & 14.86$\pm$0.02 & 14.97 \\
\hline
oblique & oblique1  
& -3.0$\pm$0.1 & 11.1$\pm$0.1 & 2.0  & 10.9 & 14.51$\pm$0.01 & 14.84 \\ 
&  
& 30.8$\pm$0.3 & 8.1$\pm$0.4 & 7.3  & 3.5 & 13.31$\pm$0.02 & \\
& oblique2  
& 1.3$\pm$0.1 & 15.0$\pm$0.1 &  2.0  & 14.9 & 14.88$\pm$0.01  & 15.24 \\
& oblique3  
& -39.5$\pm$10.2 & 59.4$\pm$7.9 &  20.1 & 55.9 & 13.69$\pm$0.05 & 15.18 \\ 
&  
& 7.1$\pm$0.1 & 16.5$\pm$0.4 & 3.2   & 16.2 & 15.04$\pm$0.03 & \\
\hline 
\multicolumn{8}{@{}l}{\parbox{0.9\textwidth}{\small\vspace{1mm}{Same as Table~\ref{tab:s1100}, but for Run 10 at 100 Myrs.}}} \\
\end{tabular}
\end{table*}

\begin{table*}[!ht]
\caption{Run 10 at 150 Myrs}
\hspace{-1.2cm}
\label{tab:my_complex_table_run10_150Myr}
\begin{tabular}{l l|*{5}{c}|c}
\hline
\multicolumn{2}{c|}{\multirow{2}{*}{Sight Lines}} & \multicolumn{5}{c|}{Results} & \multirow{2}{*}{$\log N(O~VI)_{sim}$} \\ 
\cline{3-7}
& & v$_{c}$ (km s$^{-1}$) & $b$ (km s$^{-1}$) & $b_T$ (km s$^{-1}$) & $b_{nt}$ (km s$^{-1}$) & $\log N(O~VI)$ \\ 
\hline
y & y1 
& 2.7$\pm$0.1 & 8.9$\pm$0.1 & 1.7  & 8.7 & 14.21$\pm$0.00 & 14.78 \\
& y2 
& -6.2$\pm$0.1 & 9.2$\pm$0.1 &  2.1 & 9.0 & 14.19$\pm$0.01 & 14.69 \\ 
& y3  
& -3.4$\pm$0.1 & 12.9$\pm$0.3 & 2.1  & 12.7 & 15.08$\pm$0.04 & 15.22 \\
\hline
z & 
z1  
& 10.0$\pm$1.1 & 35.1$\pm$1.2 &  3.9 & 34.9 & 15.31$\pm$0.04 & 15.97 \\ 
&   
& 11.9$\pm$2.7 & 104.2$\pm$8.3 &  23.5  & 101.5 & 14.31$\pm$0.06 &  \\ 
&   
& 54.7$\pm$1.0 & 19.7$\pm$0.9 &  3.4 & 19.4 & 15.04$\pm$0.05 &  \\ 
& z2  
& 22.2$\pm$1.4 & 23.6$\pm$2.4 &  7.3 & 22.4 & 14.82$\pm$0.21 & 14.67 \\ 
&  
& 60.6$\pm$2.6 & 29.4$\pm$3.8 & 7.7  & 28.4 & 14.61$\pm$0.06 & \\
& z3  
& 56.7$\pm$0.2  & 12.5$\pm$0.2 & 2.5  & 12.2 & 14.58$\pm$0.01 & 15.03 \\ 
& 
& 70.2$\pm$0.6 & 13.1$\pm$0.5 & 3.2 & 12.7 & 14.07$\pm$0.04 & \\
\hline
oblique & oblique1  
& -5.8$\pm$0.1 & 11.8$\pm$0.1 &  2.2 & 11.6 & 14.50$\pm$0.00 & 15.05  \\ 
&  
& 20.4$\pm$0.6 & 12.9$\pm$1.2 &  12.4 & 3.6 & 12.38$\pm$0.03 & \\
& oblique2  
& 11.3$\pm$0.2 & 11.4$\pm$0.3 &  2.2  & 11.2 & 14.42$\pm$0.01  & 14.77 \\ 
& 
& 27.2$\pm$0.6 & 7.3$\pm$0.6 & 2.6  & 6.8 & 13.51$\pm$0.06 & \\
& oblique3  
& 23.6$\pm$0.1 & 10.3$\pm$0.1 &  2.2 & 10.1 & 14.36$\pm$0.01 & 14.99 \\ 
&  
& 40.0$\pm$0.2 & 7.7$\pm$0.3 &  2.8  & 7.2 & 13.47$\pm$0.02 & \\
\hline 
\multicolumn{8}{@{}l}{\parbox{0.9\textwidth}{\small\vspace{1mm}{Same as Table~\ref{tab:s1100}, but for Run 10 at 150 Myrs.}}} \\
\end{tabular}
\end{table*}

\begin{table*}[!ht]
\caption{Run 10 at 200 Myrs}
\hspace{-1.2cm}
\label{tab:my_complex_table_run10_200Myr}
\begin{tabular}{l l|*{5}{c}|c}
\hline
\multicolumn{2}{c|}{\multirow{2}{*}{Sight Lines}} & \multicolumn{5}{c|}{Results} & \multirow{2}{*}{$\log N(O~VI)_{sim}$} \\ 
\cline{3-7}
& & v$_{c}$ (km s$^{-1}$) & $b$ (km s$^{-1}$) & $b_T$ (km s$^{-1}$) & $b_{nt}$ (km s$^{-1}$) & $\log N(O~VI)$ \\ 
\hline
y & y1 
& -7.6$\pm$0.1 & 16.9$\pm$0.1 &  6.2 & 15.7 & 13.18$\pm$0.01 & 14.98 \\ 
&   
& -2.7$\pm$0.1 & 7.8$\pm$0.1 &  2.0 & 7.5 & 14.07$\pm$0.01 &  \\ 
& y2 
& -1.8$\pm$0.1 & 12.2$\pm$0.1 & 2.0  & 12.0 & 14.52$\pm$0.01 & 15.18 \\ 
& y3  
& -2.6$\pm$0.1 & 8.4$\pm$0.1 &  2.0 & 8.2 & 14.15$\pm$0.01 & 14.87 \\
\hline
z & 
z1  
& 36.5$\pm$5.8 & 89.5$\pm$12.2 & 29.3  & 84.6 & 13.79$\pm$0.10 & 16.38 \\ 
&   
& 50.1$\pm$0.1 & 27.9$\pm$0.4 &  2.5  & 27.8 & 16.33$\pm$0.05 &  \\
& z2  
& 27.5$\pm$2.0 & 72.4$\pm$2.3 &  25.7 & 67.7 & 13.95$\pm$0.03 & 16.21 \\ 
&  
& 33.7$\pm$0.5 & 19.9$\pm$0.3 & 2.6  & 19.7 & 15.69$\pm$0.04 & \\ 
& 
& 72.1$\pm$0.3 & 16.6$\pm$0.2 & 2.4 & 16.4 & 15.18$\pm$0.02 & \\
& z3  
& 24.3$\pm$0.2  & 8.2$\pm$0.3 &  5.7 & 5.9 & 14.09$\pm$0.02 & 15.40 \\ 
& 
& 62.0$\pm$1.4 & 37.6$\pm$1.5 & 4.5 & 37.3 & 14.55$\pm$0.03 & \\ 
& 
& 76.9$\pm$0.4 & 15.0$\pm$0.7 & 4.1 & 14.4 & 14.89$\pm$0.05 & \\
\hline
oblique & oblique1  
& 10.4$\pm$0.2 & 13.3$\pm$0.4 &  2.2 & 13.1 & 14.51$\pm$0.01 & 14.61 \\ 
&  
& 31.9$\pm$1.4 & 13.5$\pm$2.2 &  4.5 & 12.7 & 13.45$\pm$0.08 & \\
& oblique2  
& 16.0$\pm$0.1 & 11.5$\pm$0.1 &  2.0  & 11.3 & 14.46$\pm$0.01  & 15.03 \\
& oblique3  
& 21.5$\pm$0.1 & 11.5$\pm$0.1 &  2.0 & 11.3 & 14.46$\pm$0.01 & 15.08 \\ 
&  
& 45.0$\pm$0.1 & 6.7$\pm$0.1 &  2.4  & 6.3 & 13.50$\pm$0.01 & \\
\hline 
\multicolumn{8}{@{}l}{\parbox{0.9\textwidth}{\small\vspace{1mm}{Same as Table~\ref{tab:s1100}, but for Run 10 at 200 Myrs.}}} \\
\end{tabular}
\end{table*}

\clearpage

\clearpage
\bibliography{LineWidthManuscript}{}
\bibliographystyle{aasjournal} 

\end{document}